\documentclass[manuscript,screen]{acmart}

\usepackage{graphicx}
\usepackage{textcomp}
\usepackage{color}
\usepackage{algorithm}

\usepackage{algpseudocode}
\floatname{algorithm}{Protocol}

\usepackage{xspace}
\usepackage{amsmath}

\newcommand{\sysname}{\textsc{FedXGB}\xspace}

\newcommand{\func}[1]{{\texttt{{#1}}}}

\usepackage{multirow}
\usepackage{array}
\usepackage{diagbox}

\usepackage{subfigure}
\usepackage{epstopdf}

\AtBeginDocument{%
  \providecommand\BibTeX{{%
    \normalfont B\kern-0.5em{\scshape i\kern-0.25em b}\kern-0.8em\TeX}}}

\setcopyright{acmcopyright}
\copyrightyear{2018}
\acmYear{2018}
\acmDOI{10.1145/1122445.1122456}

\acmConference[Woodstock '18]{ACM Transactions on Multimedia Computing, Communications, and Applications}{June 03--05, 2018}{Woodstock, NY}
\acmBooktitle{ACM Transactions on Multimedia Computing, Communications, and Applications,
  June 03--05, 2018, Woodstock, NY}
\acmPrice{15.00}
\acmISBN{978-1-4503-XXXX-X/18/06}

\begin{document}

\title{Cloud-based Federated Boosting for Mobile Crowdsensing}

\author{Zhuzhu Wang}
\email{zzwang_2@stu.xidian.edu.cn}

\author{Yilong Yang}
\authornote{Corresponding author.}
\email{echotoken@gmail.com}


\author{Yang Liu}
\email{bcds2018@foxmail.com}
\affiliation{%
  \institution{Xidian University}
  \postcode{710071}
}

\author{Ximeng Liu}
\email{snbnix@gmail.com}
\affiliation{%
  \institution{Fuzhou University}
}

\author{Brij B. Gupta}
\email{bbgupta.nitkkr@gmail.com}
\affiliation{%
  \institution{National Institute of Technology Kurukshetra}
}

\author{Jianfeng Ma}
\email{jfma@mail.xidian.edu.cn}
\affiliation{%
  \institution{Xidian University}
  \postcode{710071}
}


\renewcommand{\shortauthors}{Z. Wang and Y. Yang, et al.}

\begin{abstract}
The application of federated extreme gradient boosting to mobile crowdsensing apps brings several benefits, in particular high performance on efficiency and classification. 
However, it also brings a new challenge for data and model privacy protection. 
Besides it being vulnerable to Generative Adversarial Network (GAN) based user data reconstruction attack, there is not the existing architecture that considers how to preserve model privacy. 
In this paper, we propose a secret sharing based federated learning architecture (\textsc{FedXGB}) to achieve the privacy-preserving extreme gradient boosting for mobile crowdsensing. 
Specifically, we first build a secure classification and regression tree (CART) of XGBoost using secret sharing.
Then, we propose a secure prediction protocol to protect the model privacy of XGBoost in mobile crowdsensing. 
We conduct a comprehensive theoretical analysis and extensive experiments to evaluate the security, effectiveness, and efficiency of \textsc{FedXGB}.  
The results indicate that \textsc{FedXGB} is secure against the honest-but-curious adversaries and attains less than 1\% accuracy loss compared with the original XGBoost model.
\end{abstract}



\maketitle

\section{Introduction}
\label{Introduction}
Extreme gradient boosting (XGBoost) is an efficient, flexible and portable model that has a good performance in dealing with classification and regression, and hence applied in many apps such as malware detection~\cite{wang2017xgboost} and consumption behaviour prediction~\cite{xingfen2018research}.
Highly optimized multicore design, tactfully distributed implementation and enhanced ability to handle sparse data contribute to the success of XGBoost~\cite{chen2016xgboost}.
As a machine learning algorithm, the performance of XGBoost depends on the quality of dataset. Therefore, most companies and institutions will collect datasets with good performance by themselves but this will in need of lots of manpower~\cite{lenzen2013building} and material resources.
Hence, mobile crowdsensing (MCS), collecting data from volunteer users willing to share data, was proposed.
Recently, the privacy protection in MCS needs to be solved urgently~\cite{wang2020privacy}. 

Consider the existing mobile crowdsensing architecture, a central cloud server, owned by a service provider, collects the distributed user data and builds a machine learning model.
Such architecture suffers from two limitations: 
(1) the service provider has the heavy computational cost on the central cloud server since it not only stores a large amount of user data but also builds the machine learning model~\cite{wang2018edge};
(2) the service provider may leak the privacy because user private data are operated in the central cloud server in plaintext.
Such data leakage may cause severe problems for not only individuals but also organizations.
Nearly 3 million encrypted customer credit card records are stolen due to the data leakage of the famous company Adobe, and this caused Adobe to pay for up to \$$1$ million.

To address the above two limitations, the federated learning (FL) architecture is proposed~\cite{mcmahan2016communication}. 
FL is a kind of machine learning method that allows distributed users to not upload sensitive private data but calculated gradients~\cite{liu2020learn}.
\vspace{-0.3cm}
\begin{figure}[ht!]
\centering
\includegraphics[scale=0.95]{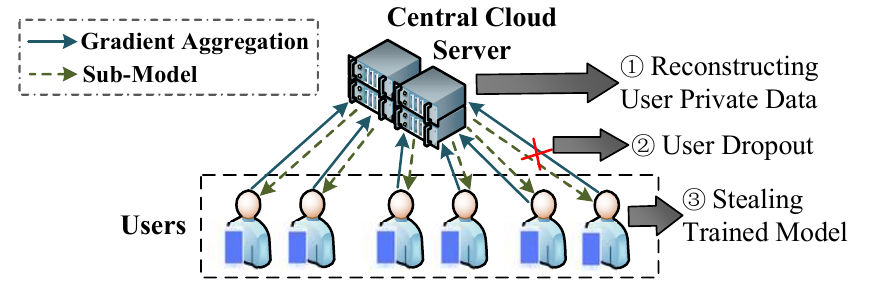}
\caption{Unattended Problems for the Current Federated Learning Architecture}
\label{problems}
\end{figure}
\vspace{-0.2cm}

Although users are protected against private data leakage, the federated learning architecture for mobile crowdsensing leads other security issues, shown in Fig.~\ref{problems}. The detailed instructions are as follows:

\begin{itemize}
    \item \textbf{User Data Reconstruction.} 
    Recent studies pointed out that the federated learning architectures are vulnerable to user data reconstruction attack~\cite{wang2019beyond},\cite{cheng2019secureboost}.
    
    A malicious central cloud server can retrieve user private data by exploiting gradient aggregation results uploaded by users based on the generative adversarial networks (GAN).
    
    \item \textbf{Model Privacy Leakage.} 
    Existing federated learning architectures built a model relying on publishing the newly built model to all users for the next round of model training~\cite{yang2019federated}.
    However, users are not always reliable.
    The trained model can potentially be stolen by adversaries with very little expense (i.e., some computation power and registration cost). 
    
    \item \textbf{User Dropout.} 
    Instability of users is not handled by the federated learning architecture~\cite{mcmahan2016communication} or its follow-up apps on specific models~\cite{zhao2018inprivate} \cite{tran2019federated}. 
    Previous architectures make an assumption that users' connectivity to the server remain steady.
    Once an user is dropped out, they have no choice but to abandon the current round of training.
    
\end{itemize}

To resolve the above issues, we propose a secret sharing based federated learning architecture (\sysname) to achieve privacy-preserving training of XGBoost for mobile crowdsensing. 
\sysname is composed of three kinds of entities, a central cloud server, edge servers and users.
Compared with the architecture shown in Fig.~\ref{problems}, edge servers are included to reflect the emerging edge computing architecture and provide a layer of privacy protection at the architectural level\cite{tong2016hierarchical}. 
\sysname proceeds in two steps.
First, it invokes a suite of secret sharing based protocols for the privacy-preserving classification and regression tree (CART) building of XGBoost.
These protocols protect user private data against the reconstruction attack.
Instead of directly publishing the newly built CART, \sysname applies a secret sharing based prediction protocol for user data updating. 
It encrypts the model from being stolen by users.

In summary, our main contributions can be summarised as follows:
\begin{itemize}
    \item \textbf{Boosting Privacy.}
    \sysname utilizes the secret sharing technique to protect user private data and the updated model during each round of federated XGBoost training. 
    The gradients of user private data are secretly shared to protect them against the user data reconstruction attacks. 
    The model prediction are performed with secretly shared parameters to address the model privacy leakage. 
    
    \item \textbf{Low Accuracy Loss.}
    We evaluate \sysname by applying two popular datasets. 
    The results indicate that \sysname maintains the high performance of XGBoost with less than 1\% of accuracy loss.

    \item \textbf{Robust Architecture against User Dropout.}
    We validate that \sysname stays stable to execute each round of training when the user dropout happens. 
    The effectiveness and efficiency are barely affected by the dropout of users. 
    
\end{itemize}

The rest of this paper is organized as follows. 
In Section \ref{Premilinary}, we briefly introduce some background knowledge.
Section \ref{SystemArchitecture} describes an overview of \sysname. 
Section \ref{Main_Approach} and Section \ref{SecurePrediction} list the implementation details of \sysname. 
Section \ref{SecurityAnalysis} discusses the security of \sysname.
In Section \ref{Experiments}, we perform a series of comprehensive experiments.
Section \ref{RelatedWork} discusses the related work. 
The last section concludes the paper.

\section{PRELIMINARY}
\label{Premilinary}
The background knowledge about XGBoost, secret sharing and two cryptographic functions are briefly introduced in this section.
For convenience, notations used in this paper are summarized in Table~\ref{table_notation}.
\vspace{-0.2cm}
\renewcommand\arraystretch{1.3}
\begin{table}[!htbp]
\centering
\caption{Notations}
\begin{tabular}{cl}

\hline
Notation & Description       \\


\hline

$l(\cdot)$          & an arbitrary loss function with second-order derivative\\

$g_i$               &  the first-order derivative of $l(\cdot)$ for the $i_{th}$ instance \\

$h_i$               &  the second-order derivative of $l(\cdot)$ for the $i_{th}$ instance \\

$\zeta_u$           & the secret share distributed to the user $u$  \\

$\mathcal{R}_u$     & the set of random share of private mask key for user $u$ \\

$\mathcal{F}$       & a finite field $\mathcal{F}$, e.g. $\mathcal{F}_p=\mathcal{Z}_p$ for some large prime $p$ \\

$f_k$               & the CART obtained from the $k$-th iteration of XGBoost\\

$\langle\cdot\rangle_u$        & key for signature, encryption or secret mask generation\\

\hline
\end{tabular}
\label{table_notation}
\end{table}
\vspace{-0.4cm}

\subsection{Extreme Gradient Boosting}\label{pre_XGBoost}
XGBoost is one of the most outstanding ensemble learning methods due to its excellent performance in processing classification, regression, and Kaggle tasks\cite{mitchell2017accelerating}, which implements machine learning algorithms under the Gradient Boosting framework by establishing numerous classification and regression trees (CART) models.
The core of algorithm is the optimization of the value of the objective function $\mathcal{L}^{k}$ as follows.
\begin{equation}\label{XGB_OBJ}
\begin{aligned}
    \mathcal{L}^{k} = \sum^n_{i=1} l(y_i, \hat{y}_i^{k-1} + f_k(x_i)) + \Omega (f_k),\\
\end{aligned}
\end{equation}
where $k$ is the number of iterations, $n$ is the total number of training samples, $i$ is the index of each sample, and $y_i$ is the label of the $i$-th sample.
$\hat{y}_i^{k-1}$ represents the predicted label of the $i$-th sample at the ($k - 1$)-th iteration.
$\Omega$ is a regularization item.
By expanding $\Omega$ and using the second-order Taylor approximation,
    the optimal weight $\omega_j$ of leaf node $j$ is calculated as follows.
\begin{equation}\label{XGB_OPT}
    \omega_j = -\sum^{T}_{j = 1} \frac{(\sum_{i\in I_j} g_i)^2}{\sum_{i\in I_j} h_i + \lambda},
\end{equation}
where $I_j$ represents the set of training samples in the leaf node.
$\lambda$ is a constant value.
$T$ is the number of tree leaves.
According to the above equations, 
    we can retrieve an optimal tree for the $k$-th iteration.
\vspace{-0.2cm}

\subsection{Secret Sharing}
Since attackers can easily derive user private data by exploiting the uploaded gradients~\cite{cheng2019secureboost}, 
    the $(t,n)$ Secret Sharing ({SS}) scheme~\cite{shamir1979share} is adopted in our scheme.
For the $(t, n)$ {SS} scheme, 
    a secret $s$ is split into $n$ shares.
$s$ is recovered only if at least $t$ random shares are provided;
    otherwise, it cannot be obtained.
The share generation algorithm is illustrated as \texttt{SS.share}$(s, t, n) = \{(u, \zeta_u)| u\in \mathcal{U} \}$,
    in which $n$ represents the number of users involved in {SS} and $\mathcal{U} = \{1, 2, ,...,n\}$ is a set including these users.
$\zeta_u$ describes the share for each user $u$.
To recover the secret,
    the Lagrange polynomials based recovery algorithm \texttt{SS.recon}$(\{(u, [\zeta]_u)| u\in \mathcal{U}' \}, t)$ is used.
It requires that $\mathcal{U}'\subseteq \mathcal{U}$ has to contain at least $t$ users.

We apply a secret sharing based comparison protocol (SecCmp)~\cite{huang2019lightweight} to fulfill the secure comparison in \sysname. 
Without revealing the plaintext values to edge servers, \func{SecCmp} returns the comparison result to the user.

\textbf{Secure Comparison Protocol (SecCmp).}
Given two sets of secret shares, 
        \texttt{SS.Share}$(s_1, t, n) = \{(u, \zeta_{u}^1)| u\in \mathcal{U} \}$ and \texttt{SS.Share}$(s_2, t, n) = \{(u, \zeta_{u}^2)| u\in \mathcal{U} \}$, 
            the random shares of the comparison result $\{(u, \zeta_{u})| u\in \mathcal{U}\}$ is generated. 
Having at least $t$ shares, i.e., $n\geq |\mathcal{U}'| > t$ and $\mathcal{U}' \subset \mathcal{U}$,
    the result is recovered.
If $s_1 > s_2$, \texttt{SS.Recon}$(\{(u, \zeta_u)| u\in \mathcal{U}' \}, t) = 0$;
otherwise,
    \texttt{SS.Recon}$(\{(u, \zeta_u)| u\in \mathcal{U}' \}, t) = 1$.
\vspace{-0.2cm}

\subsection{Cryptographic Definition}\label{sec_crypto}
To securely transmit data, 
    three cryptographic functions are utilized in \sysname.

\subsubsection{Key Agreement}
Key agreement is used for key generation.
Three algorithms are involved for key agreement, namely key setup \texttt{KEY.Set}, key generation \texttt{KEY.Gen}, and key agreement \texttt{KEY.Agr}.
Specifically, the key setup algorithm, \texttt{KEY.Set}$(\ell)$, is for setting up a public parameter $p_{pub}$. $\ell$ is a security parameter that defines the field size of a secret sharing scheme $\mathcal{F}_p$. \texttt{KEY.Set}$(\ell)$ outputs a quaternion $p_{pub}\gets(\mathbb{G}, p, g, H)$.
$\mathbb{G}$ is an additive cyclic group with a large prime order $p$ and a generator $g$. And $H$ is a common hash function that generates a fixed length output. 
Consider two arbitrary users, $u$ and $v$, $u$ first applies the key generation algorithm to generate a private-public key pair $(\langle k \rangle^{pri}_u, \langle k \rangle^{pub}_u) = $\texttt{KEY.Gen}$(p_{pub})$.
Then, $u$ can use the key agreement algorithm to create a shared key with $v$, $\langle k \rangle_{u, v}\gets$\texttt{KEY.Agr}$(\langle k \rangle^{pri}_u,\langle k \rangle^{pub}_v)$.

\subsubsection{Identity Based Encryption \& Signature}
Identity based encryption and signature are utilized to encrypt sensitive data and verify identity, respectively.
Given a shared key $\langle ek \rangle^{enc} = \langle ek \rangle^{dec}\gets$\texttt{KEY.Agr}, the identity based encryption algorithm \texttt{IDE.Enc} outputs ciphertext $c = $\texttt{IDE.Enc}$(\langle ek \rangle^{enc}, t)$.
And the decryption function \texttt{IDE.Dec} recovers the plaintext $t$ by computing $t = $\texttt{IDE.Dec}$(\langle ek \rangle^{dec}, c)$.
Similarly,
    the signature algorithms, \texttt{SIG.Sign} and \texttt{SIG.Verf}, are defined.
Given the key pair for signature $(\langle k \rangle^{sig}, \langle k \rangle^{ver})\gets$\texttt{KEY.Gen}, \texttt{SIG.Sign} outputs a signature $\sigma =  $\texttt{SIG.Sign} $(\langle k \rangle^{sig}, t)$.
If \texttt{SIG.Verf}$(\langle k \rangle^{ver}, t, \sigma) = 1$, $\sigma$ is proved to be valid; otherwise, $\sigma$ is invalid.

\section{Overview of \sysname}
\label{SystemArchitecture}
In this section, we introduce how the secret sharing based federated learning architecture (\sysname) is implemented. 
\vspace{-0.2cm}

\subsection{Entities of \sysname}
\begin{figure}[ht!]
\centering
\includegraphics[scale=0.75]{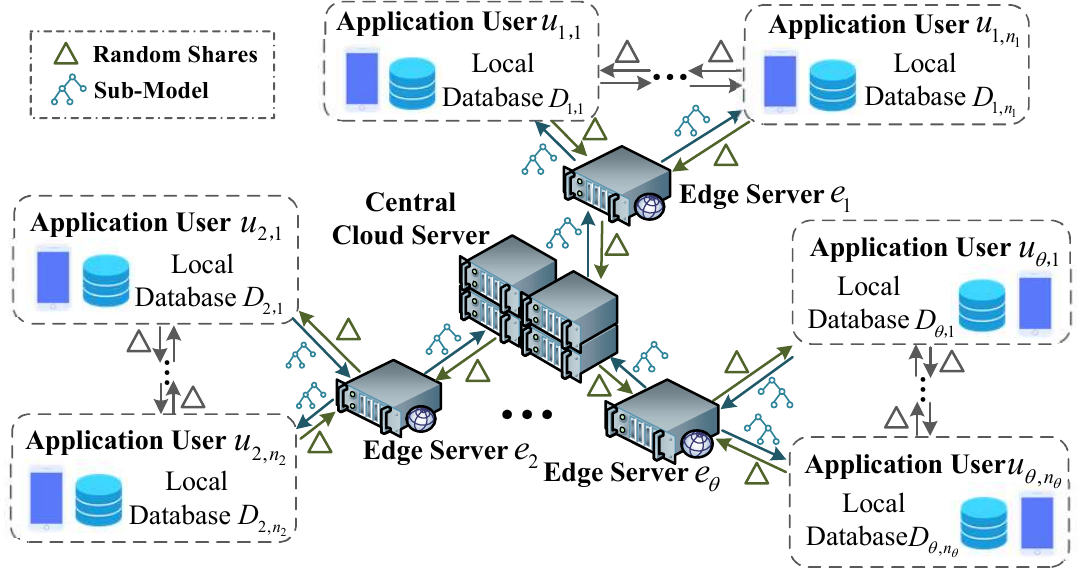}
\caption{Entities of \sysname}
\label{system_model}
\end{figure}

\sysname consists of three types of entities: users $\mathcal{U}$, edge servers $\mathcal{E}$, and a remote central cloud server $\mathcal{S}$. The entities of \sysname are showed in Fig.~\ref{system_model}.
Details are presented as follows.
\vspace{-0.2cm}
\vspace{-0.2cm}

\textbf{Users.} 
$\mathcal{U} = \{\mathcal{U}_1, \mathcal{U}_2, ..., \mathcal{U}_{\theta}\}$. For each $\mathcal{U}_i \in \mathcal{U}$, $\mathcal{U}_i = \{u_{i,1}, u_{i,2}, ..., u_{i,n_i}\}$ represents a set of users belonging to the domain $i$. 
Users are data generators and volunteers to participate in the crowdsensing model training for \sysname.

\textbf{Edge Servers.} 
$\mathcal{E} = \{e_1, e_2, ..., e_{\theta}\}$, where $e_j\in \mathcal{E}$ is an edge server. 
Edge servers are provided by various operators.
Each edge server provides the communication service for users that belong to the domain it controls.

\textbf{Central Cloud Server.} 
$\mathcal{S}$ is a central cloud computing server owned by a mobile crowdsensing service provider.
The trained model in \sysname only belongs to $\mathcal{S}$, and is not publicly accessible.
\vspace{-0.2cm}

\subsection{Security Model}
In \sysname, we use the \textit{curious-but-honest} model as our standard security model. The definition of the adversary $\mathcal{A}$ in our security model is formalized as follows:

\noindent 
\textbf{Definition 1~\cite{paverd2014modelling}.}
\textit{In a communication protocol, a legitimate entity, $\mathcal{A}$, does not deviate from the defined protocol, but attempts to learn all possible information from the legitimately received messages.} 

Any $u\in \mathcal{U}$, $e\in\mathcal{E}$ and $\mathcal{S}$ can be an $\mathcal{A}$, with the following abilities: 1) $\mathcal{A}$ can corrupt or collude with less than $t$ legitimate users or edge servers and get the corresponding inputs; 2) $\mathcal{A}$ cannot extract the information from other good parties (e.g., legitimate inputs, random seeds); 3) $\mathcal{A}$ has limited computing power to launch attacks (i.e., polynomial-time attacks).
\sysname needs to achieve the following two goals.
\begin{itemize}
    \item \textbf{Data Privacy.} $e\in\mathcal{E}$ and $\mathcal{S}$ are unable to learn the private data of $u\in \mathcal{U}$, especially through the data reconstruction.
    \item \textbf{Model Privacy.} $u\in \mathcal{U}$ and $e\in\mathcal{E}$ are unable to learn the key model parameters owned by $\mathcal{S}$. 
\end{itemize}

\subsection{Workflow of \sysname}
Two protocols are involved in each round of \sysname, secure CART model building (SecBoost) and secure CART model prediction (SecPred), shown in Fig.~\ref{system_workflow}. Working details of the protocols are given below.
\begin{figure}[ht!]
\centering
\includegraphics[scale=0.6]{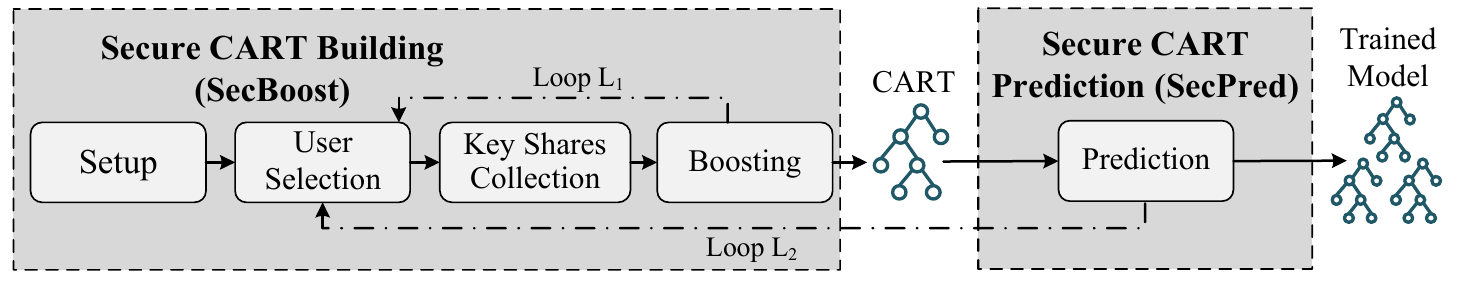}
\caption{Workflow of \sysname}
\label{system_workflow}
\end{figure}

\func{SecBoost} takes the following four steps:
 
\begin{enumerate}
    \item \textbf{Setup.} 
    All entities (i.e., $\mathcal{U}$, $\mathcal{E}$, and $\mathcal{S}$) setup essential parameters for CART model building, including preset parameters of XGBoost and cryptographic keys.
    
    \item \textbf{User Selection.} 
    According to the predefined standards, each edge server selects the active users in its domain.
    The selected users verify the identity with each other and exchange the public keys.
    Additionally, each selected user creates a mask key pair for further secret sharing between the user and its corresponding edge server.
    
    \item \textbf{Key Shares Collection.}
    Depending on the number of selected users, each user generates random shares of the private mask key and sends each share to the other selected users.
    The set of key shares is constructed by collecting the random shares from other users, and used to recover the private mask key when the user drops out.
   
    \item \textbf{Boosting.}
    To build a CART model securely, the selected users mask the locally calculated sub-aggregation of gradients and upload the masked value to edge servers. 
    Then, the edge servers sum the masked sub-aggregations and send the results to the central cloud server.
    The central server adds the received values up for further CART model building.
    \sysname iteratively calculates the optimal split to extract an optimal CART model until the termination condition is met (Loop L$_1$ in Fig.~\ref{system_workflow}).
    
\end{enumerate}

\func{SecPred} is designed to extract prediction results of the newly obtained CART without model privacy disclosure.
$\mathcal{S}$ executes \func{SecPred} by taking the following steps: 
    a) all splits of the CART model calculated by $\mathcal{S}$ and user private data are secretly shared to $\mathcal{E}$;
    b) $\mathcal{E}$ repeatedly invokes \func{SecCmp} to compute the comparison results;
    c) The comparison results are sent to $\mathcal{U}$ for updating $\hat{y}_i^{k-1}$, mentioned in Eq.~\ref{XGB_OBJ}.

After \func{SecPred} terminates, \sysname completes one round of training and begins the next round of training (i.e., Loop L$_2$ shown in Fig.~\ref{system_workflow}).
When L$_2$ is completed, $\mathcal{S}$ gets a trained XGBoost model $ f(x) = \sum_{\kappa  = 1}^K f_{\kappa}(x)$, where $K$ is the maximum training round.

The security goals of \sysname are achieved as follows. 
For the first goal, the gradient sub-aggregation of users are protected with the secret sharing technique in \func{SecBoost} to defend the reconstruction attack proposed in \cite{cheng2019secureboost}.
And the tree structure of XGBoost is chosen against the other type of reconstruction attack proposed in \cite{wang2019beyond}.
The security analysis for the goal is given in Section~\ref{SecurityAnalysis} and Section~\ref{sec_UDR}.
The second goal is also achieved by the secret sharing technique in \func{SecPred}, whose security analysis is discussed in Section~\ref{SecurityAnalysis}.
\vspace{-0.2cm}
  
\section{Secure CART building of \sysname}
\label{Main_Approach}
In this section, we present the protocol \func{SecBoost} in details.

In All users in \sysname are orderly labeled by a sequence of indexes (i.e., $1, 2, ..., n$) to represent their identities. 
Each user is deployed a small local dataset $D_u$.
The message $m$ sent from $A$ to $B$ is briefly described as $A\Rightarrow B$ : $m$.

\subsection{Secure CART Model Building}
\begin{figure}[ht!]
\centering
\includegraphics[scale=0.75]{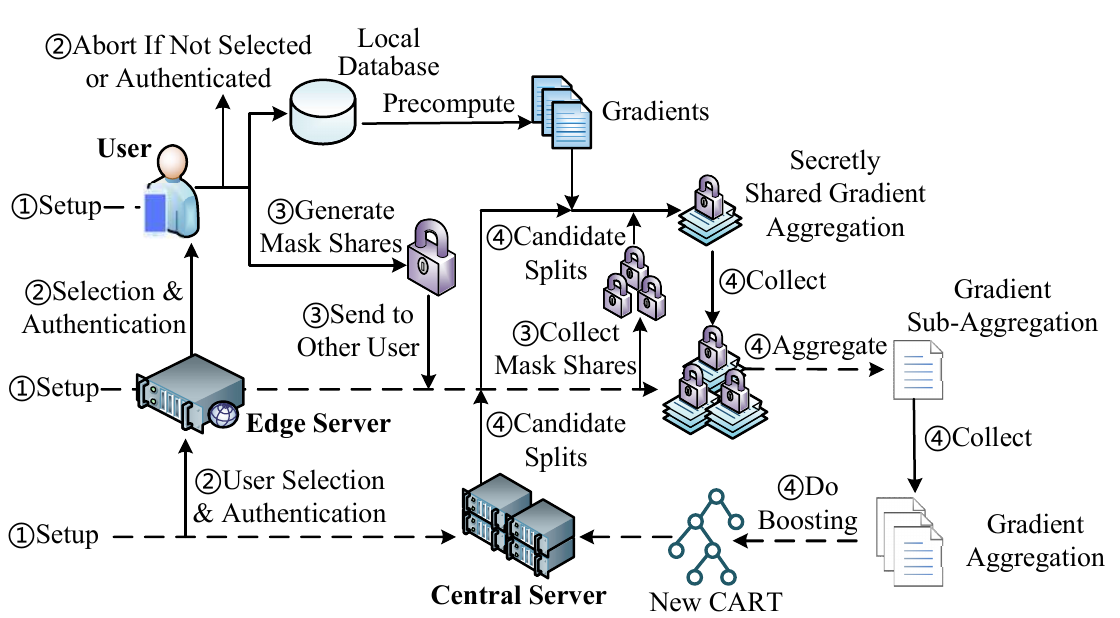}
\caption{High-level Overview of \texttt{SecBoost}}
\label{overflow}
\end{figure}

Protocol \func{SecBoost} is for secure CART model building, shown in \textbf{Protocol \ref{FedLearning}}.
Refer to the overview of \func{SecBoost} illustrated in Fig.~\ref{overflow}, we introduce the steps in details as below.

\textbf{Step 1 - Setup:} 
$\mathcal{U}$, $\mathcal{E}$, and $\mathcal{S}$ setup public parameters for key generation and model training.
Firstly, \sysname is given 
    the input space field $\mathcal{F}_q$, 
    the secret sharing field $\mathcal{F}_p$,
    the parameter for key generation $p_{pub}=$\texttt{KEY.Set}$(\ell)$,
    and the publicly known {XGBoost} parameters including $\lambda$, $l(\cdot)$ and maximum tree depth $d_{max}$.
    and the publicly known {XGBoost} parameters including $\lambda$, $\gamma$, $l(\cdot)$ and maximum tree depth $d_{max}$.
Then,
    a trusted third party $\mathcal{T}$ generates a signature key pair $(\langle k \rangle^{sig}_u , \langle k \rangle^{ver}_u)$ for $u\in \mathcal{U}$.
The encryption key pair $(\langle ek \rangle^{pri}_u, \langle ek \rangle^{pub}_u)$ is generated by $u\in \mathcal{U}$.

\textbf{Step 2 - User Selection: }
To minimize the cost for data recovery of dropout users,
    $e_j\in \mathcal{E}$ selects the more active users to participate in the model training.
The predefined selection standards include the keeping active time, the connection stability, and the maximum of users.
The selected users are expressed as $\mathcal{U}'\subset \mathcal{U}$.
The number of users control by $e_j$ is $n_j$, and their secret sharing threshold is $t_j$.
Through \func{KEY.Gen}$(p_{pub})$, $u\in \mathcal{U}$ generates the mask key pair $(\langle sk \rangle^{pri}_u, \langle sk \rangle^{pub}_u)$ for secret sharing.
The message for public key exchange ($\langle sk \rangle^{pub}_u$ and $\langle ek \rangle^{pub}_u$) is signed.
The legitimacy of selected users is confirmed by verifying their signatures.

\textbf{Step 3 - Key Shares Collection: }
Users generate random shares of their private mask keys by computing $\{(u, \zeta_{u, v}^{sk})| v\in \mathcal{U}_j'\} \gets$\texttt{SS.share}$(\langle sk \rangle^{pri}_u, t, n)$, and send $\zeta_{u, v}^{sk}$ to each of the other selected users $v\in\mathcal{U}_j'$ in the encrypted format.
User $v$ decrypts $c_{u, v}$ to extract $\zeta_{u, v}^{sk}$ and expands its key share set $\mathcal{R}_v = \mathcal{R}_v\cup \{(u, \zeta_{u, v}^{sk})\}$.
$\mathcal{R}_v$ is used to recover the private mask key when the user drops out, as discussed in Section~\ref{sec_UserDrop}.
The encryption key $\langle ek \rangle_{u, v}$ between $u$ and $v$ is calculated by \texttt{KEY.Agr}$(\langle ek \rangle^{pri}_u,$ $ \langle ek \rangle^{pub}_v)$.

\textbf{Step 4 - Boosting: }
Assume that the feature set of the user data is $\mathcal{Q} = \{\alpha_1, \alpha_2, ..., \alpha_q \}$. 
For boosting, $\mathcal{S}$ randomly selects a sub-sample $\mathcal{Q}'\subset \mathcal{Q}$ and invokes the secure split finding protocol (\func{SecFind}), introduced in Section~\ref{SplitFinding}, to find the optimal split.
To build a new CART model with an optimal structure, $\mathcal{S}$ successively operates the boosting process until the current tree depth reaches $d_{max}$ or other termination conditions~\cite{chen2016xgboost} are met.
Finally, \texttt{SecBoost} outputs a well trained CART model $f_{\kappa}$.
\begin{algorithm}[ht!]
  \caption{Secure Extreme Gradient Boosting Based Tree Building (\func{SecBoost})}
  \label{FedLearning}
  \begin{algorithmic}[1]
    \Require
      A central server $\mathcal{S}$, a edge server set $\mathcal{E} = \{e_1, ..., e_{\theta}\}$, a user set $\mathcal{U} = \{\mathcal{U}_1, ..., \mathcal{U}_{\theta}\}$ and a trusted third party $\mathcal{T}$.
    \Ensure
      A well-trained CART.

    \State \textbf{Step 1:}
     $\mathcal{S}$ selects security parameter $p_{pub}\gets$\texttt{KEY.Set}$(\ell)$ and publishes the model parameters $p_{pub}$, $\lambda$, $\gamma$, $l(\cdot)$, $d_{max}$.
    \State $\mathcal{T}$ generates signature key pair $(\langle k \rangle^{sig}_u, \langle k \rangle^{ver}_u)$ for $u\in\mathcal{U}$, and operates $\mathcal{T}\Rightarrow \mathcal{U}$ : $(\langle k \rangle^{sig}_u, \langle k \rangle^{ver}_u)$.
    \State $u\in \mathcal{U}_j'$ invokes $(\langle ek \rangle^{pri}_u, \langle ek \rangle^{pub}_u) \gets $\texttt{KEY.Gen}$(p_{pub})$.
    
    \State \textbf{Step 2:}
     $e_j\in \mathcal{E}$ selects a set of active users $\mathcal{U}_j'$, secret sharing threshold $t_j$ and operates $e_j\Rightarrow \mathcal{U}_j'$ : $(\mathcal{U}_j' , t_j)$.
    \State $u\in \mathcal{U}_j'$ invokes $(\langle sk \rangle^{pri}_u, \langle sk \rangle^{pub}_u)\gets$ \texttt{KEY.Gen}$(p_{pub})$.
    \State $u \Rightarrow e_j \Rightarrow \mathcal{S}$: $(u, \langle sk \rangle^{pub}_u, \langle ek \rangle^{pub}_u, \sigma_u \gets$\texttt{SIG.Sign}$(\langle k \rangle^{sig}_u,$ $\langle sk \rangle^{pub}_u || \langle ek \rangle^{pub}_u))$.
    \State $\mathcal{E}$ and $\mathcal{S}$ verify \texttt{SIG.Verf}$(k_u^{ver}, \langle sk \rangle^{pub}_u || \langle ek \rangle^{pub}_u, \sigma_u) = 1$ and forward $e_j\Rightarrow \mathcal{U}_j'$ : $(u, \langle sk \rangle^{pub}_u, \langle ek \rangle^{pub}_u, \sigma_u)$.
    \State Other users in $\mathcal{U}_j'$ verify whether \texttt{SIG.Verf} $(k_u^{ver},$ $\langle sk \rangle^{pub}_u || \langle ek \rangle^{pub}_u, \sigma_u) = 1$.
    
    \State \textbf{Step 3:}
     $u\in \mathcal{U}_j'$ computes and collects the shares of mask key $\langle sk \rangle^{pri}_u$ by invoking $\mathcal{R}_u = \{(u, \zeta_{u, v}^{sk})| v\in \mathcal{U}_j'\} \gets$ \texttt{SS.share}$(\langle sk \rangle^{pri}_u, t, n)$.
     \State $u\Rightarrow e_j\Rightarrow v$ : $c_{u, v} \gets $\texttt{IDE.Enc}$(\langle ek \rangle_{u, v}, u||v||\zeta_{u, v}^{sk})$.
     
    \State $v\in \mathcal{U}_j'$ decrypts $\zeta_{u, v}^{sk}\gets$\texttt{IDE.Dec}$(\langle ek \rangle_{u, v}, c_{u,v})$, and collects $\mathcal{R}_v = \mathcal{R}_v \cup \{(u, \zeta_{u, v}^{sk})\}_{u\in\mathcal{U}_j'}$.
    
    \State \textbf{Step 4:}
     $\mathcal{S}$ randomly selects a feature sub-sample $\mathcal{Q}'$ from full feature set $\mathcal{Q}$. 
    \State $\mathcal{S}$ invokes \texttt{SecFind}$(\mathcal{Q}', \mathcal{U}', \mathcal{E})$ to determine the current optimal split. 
    \State Repeat \textbf{Step 2} until reaching the termination condition. 
  \end{algorithmic}
\end{algorithm} 
\vspace{-0.6cm}

\subsection{Secure Split Finding for \func{SecBoost}}\label{SplitFinding}
\begin{algorithm*}[ht!]
  \caption{Secure Split Finding (\texttt{SecFind})}
  \label{SecFind}
  \begin{algorithmic}[1]
    \Require
      All candidate features $\mathcal{Q}'$, the active user set $\mathcal{U}'$, the edge server set $\mathcal{E}$.
    \Ensure
      The optimal split for feature $\alpha\in \mathcal{Q}'$ and its score.
    
    \For {$1 \leq j \leq \theta$}
        \State Each $u\in \mathcal{U}_j'$ generates a random value $r_u$ and its random shares $\{(u, \zeta_{u, v}^{r})| v\in \mathcal{U}_j'\} \gets$\texttt{SS.share}$(r_u, t, n)$. 
        \State $u\Rightarrow e_j\Rightarrow v$ : $c_{u, v} \gets $\texttt{IDE.Enc}$(\langle ek \rangle_{u, v},$ $ u||v||\zeta_{u, v}^{r})$.
        \State Each $v\in \mathcal{U}_j'$ receives $c_{u, v}$ and decrypts $\zeta_{u, v}^{r} \gets $\texttt{IDE.Dec}$(\langle ek \rangle_{u, v}, u||v||\zeta_{u, v}^{r})$.
    \EndFor
    
    \For {$1 \leq j \leq \theta$}
        \State $u\in \mathcal{U}_j'$ generates $\langle sk \rangle_{u, v} = $\texttt{KEY.Agr}$(\langle sk \rangle_{u}^{pri}, \langle sk \rangle_{v}^{pub})$ for each $v\in \mathcal{U}_j'$.
        \State $u\in \mathcal{U}_j'$ computes $\zeta_u^{H}\gets$\texttt{SecMask}$(\sum_{1\leq k \leq |D_u|} h_k, r_u,  \langle sk \rangle_{u, v})$ and $\zeta_u^{G}\gets$\texttt{SecMask}$(\sum_{1\leq k \leq |D_u|} g_k, r_u,  \langle sk \rangle_{u, v})$. 
        \State $u\Rightarrow e_j$ : $c_{u, e_j}\gets$\texttt{IDE.Enc}$(\langle ek \rangle_{u, e_j}, \zeta_u^{H} || \zeta_u^{G} ||$ $ \{\zeta_{v, u}^{r}\}_{v\in \mathcal{U}_j'})$.
        \State $e_j$ decrypts $c_{u, e_j}$ and reconstructs $r_u = $\texttt{SS.Recon}$($ $\{\zeta_{u, v}^{r}\}_{v\in \mathcal{U}_j'}, t)$.
        \State $e_j\Rightarrow \mathcal{S}$ : $H_{j} \gets$ \texttt{IDE.Enc}$(\langle ek \rangle_{e_j, \mathcal{S}}, \sum_{u\in \mathcal{U}_j'} (\zeta_u^{H} - r_u))$,  $G_{j} \gets$ \texttt{IDE.Enc}$(\langle ek \rangle_{e_j, \mathcal{S}}, \sum_{u \in \mathcal{U}_j'} (\zeta_u^{G} - r_u))$.
    \EndFor
    \State $\mathcal{S}$ calculates $H \gets \sum_{j=1}^{\theta}$\texttt{IDE.Dec}$(\langle ek \rangle_{e_j, \mathcal{S}},H_j)$ and $G \gets \sum_{j=1}^{\theta}$\texttt{IDE.Dec}$(\langle ek \rangle_{e_j, \mathcal{S}},G_j)$.
    
    \For {$1 \leq q \leq \delta$}
        \State $\mathcal{S}$ enumerates every possible candidate split $A_q = \{a_1, a_2, ..., a_m\}$ for feature $\alpha_q\in \mathcal{Q}'$ and publishes them to each user $u\in \mathcal{U}'$ through $\mathcal{E}$. For each $a_r\in A_q$, take the following steps.
        \State $u\in \mathcal{U}_j'$ generates a new random value $r_u$, and shares it like what it does in the first loop.
        \State $u\in \mathcal{U}_j'$ computes $\zeta_u^{H,L}\gets$\texttt{SecMask}$(\sum_{x_{u, l}< a_r} h_{l}, r_u,  \langle sk \rangle_{u, v})$ and $\zeta_u^{G,L}\gets$\texttt{SecMask}$(\sum_{x_{u, l}< a_r} g_{l}, r_u,  \langle sk \rangle_{u, v})$. 
        \State $u\Rightarrow e_j$ : $c_{u, e_j}\gets$\texttt{IDE.Enc}$(\langle ek \rangle_{u, e_j}, \zeta_u^{H, L} || \zeta_u^{G, L} || \{\zeta_{v, u}^r\}_{v\in \mathcal{U}_j'})$.
        \State $e_j$ decrypts $c_{u, e_j}$ and reconstructs $r_u = $\texttt{SS.Recon}$($ $\{\zeta_{u, v}^{r}\}_{v\in \mathcal{U}_j'}, t)$.
        \State $e_j\Rightarrow \mathcal{S}$ : $H_{j}^L \gets$ \texttt{IDE.Enc}$(\langle ek \rangle_{e_j, \mathcal{S}}, \sum_{u\in \mathcal{U}_j'} (\zeta_u^{H, L} - r_u))$,  $G_{j}^L \gets$ \texttt{IDE.Enc}$(\langle ek \rangle_{e_j, \mathcal{S}}, \sum_{u \in \mathcal{U}_j'} (\zeta_u^{G, L} - r_u))$.
            \State $\mathcal{S}$: $H_L \gets \sum_{j=1}^{\theta}$\texttt{IDE.Dec}$(\langle ek \rangle_{e_j, \mathcal{S}}, H_j^L)$, $H_R = H - H_L$ and $G_L \gets \sum_{j=1}^{\theta}$\texttt{IDE.Dec}$(\langle ek \rangle_{e_j, \mathcal{S}}, G_j^L)$, $G_R = G - G_L$.
            
            \State $\mathcal{S}$ then obtains $score\gets \max(score, \frac{G_L^2}{H_L + \lambda} + \frac{G_R^2}{H_R + \lambda} - \frac{G^2}{H + \lambda})$. 
    \EndFor
  \end{algorithmic}
\end{algorithm*}

The most important operation in XGBoost training is to optimize the tree structure by finding the optimal split for each node of the CART. 
In \sysname, we propose a novel secret sharing based split finding protocol \func{SecFind}, presented in \textbf{Protocol~\ref{SecFind}}.
Details of \func{SecFind} are as follows.
    
First, each user $u\in \mathcal{U}_j'$ generates a random value $r_u$ for masking secret.
Each random share $\zeta^{r}_{u, v}$ of $r_u$ is distributed to a specific user $v$.
The secret masking function (SecMask) is given as follows.
\begin{equation}\label{eq_SecretMask}
    \texttt{SecMask}(s, r_u,  \langle sk \rangle_{u, v}) =
    s + r_u + \sum_{u > v} \langle sk \rangle_{u, v} - \sum_{u < v} \langle sk \rangle_{u, v},
\end{equation}
where $s$ is a secret value, $v\in\mathcal{U}_j'$, $\langle sk \rangle_{u, v}=$\texttt{KEY.Agr}$($ $\langle sk \rangle_{u}^{pri}, \langle sk \rangle_{v}^{pub})$.
From Eq.~\ref{eq_SecretMask}, it indicates that $e_j$ can directly get $s$ when $\langle sk \rangle_{u}^{pri}$ is recovered without giving $r_u$.
As discussed in Section~\ref{sec_UserDrop}, the recovery always occurs when $u$ drops out.
Therefore, $r_u$ is an essential value for the security of \texttt{SecFind}.
The correctness of Eq.~\ref{eq_SecretMask} is given in \cite{bonawitz2017practical}.

Then,
$u$ uploads sub-aggregations of gradients for all its data, $\sum_{1\leq k \leq |D_u|} g_k$ and $\sum_{1\leq k \leq |D_u|} h_k$, mentioned in Eq.~\ref{XGB_OPT}. 
Each sub-aggregation is masked based on the Eq.~\ref{eq_SecretMask}.
$e_j$ sums all masked sub-aggregations and sends the result to $\mathcal{S}$ in the encrypted format.
In order to get the correct summing result, 
    $e_j$ has to reconstruct $r_u$ and subtracts it from the masked sub-aggregations, $\zeta_u^{H}$ and $\zeta_u^{G}$.
The encryption keys utilized here are $\langle ek \rangle_{u, e_j} = $\texttt{KEY.Gen}$(\langle ek \rangle_{u}^{pri}, \langle ek \rangle_{e_j}^{pub})$ and $\langle ek \rangle_{e_j, \mathcal{S}} = $\texttt{KEY.Gen}$(\langle ek \rangle_{e_j}^{pri}, \langle ek \rangle_{\mathcal{S}}^{pub})$
Having the summation values from each $e_j\in \mathcal{E}$, $\mathcal{S}$ adds them up to get the final aggregation result, $H$ and $G$, for all data.

Finally, for each given candidate feature $\alpha_q$, $\mathcal{S}$ enumerates all possible candidate splits and publishes them to $\mathcal{U}$. Similar to the above aggregation process for $H$ and $G$, $\mathcal{S}$ iteratively collects the left-child gradient aggregation results for each candidate split.
The aggregation results are used to compute the score for each candidate split according to Eq.~\ref{XGB_SCORE}.
When the iteration is terminated, \texttt{SecFind} outputs the split with maximum score.
Moreover, the weights of left and right child nodes about the optimal split are also determined by Eq.~\ref{XGB_OPT}.
\begin{equation}\label{XGB_SCORE}
    score = \frac{G_L^2}{H_L + \lambda} + \frac{G_R^2}{H_R + \lambda} - \frac{G^2}{H + \lambda}.
\end{equation}

\section{Secure CART Model Prediction of \sysname}\label{SecurePrediction}
In this section, the protocol \func{SecPred} is presented in details.
Besides, we discuss its robustness against user dropout.
\vspace{-0.2cm}

\subsection{Secure CART Model Prediction}
For existing federated learning schemes, an indispensable operation is to update each user's local model at the end of each round of training \cite{mcmahan2016communication}.
For XGBoost, the updated model is used for extracting prediction results to update the $\hat{y}_i^{k-1}$ in Eq.~\ref{XGB_OBJ} that are taken as input of the next round of training.
However, users are honest-but-curious entities.
They potentially steal the model information to benefit themselves (e.g. sell the model to the competitors of $\mathcal{S}$).
To protect the model privacy, \sysname executes a lightweight secret sharing protocol \texttt{SecPred}, presented in \textbf{Protocol~\ref{SecPred}}, instead of transmitting the updated {CART} model in plaintext. In \func{SecPred}, $\mathcal{S}$ takes a CART model as input and $\mathcal{U}$ takes as input of the weights of leaf nodes in the CART model.
$\mathcal{S}$ and $\mathcal{U}$ secretly and separately send the shared model parameters (optimal split to each node) and user data (feature values to each optimal split) to edge servers.
Then, $\mathcal{E}$ executes \func{SecCmp} and returns each comparison result to the corresponding user.
Finally, $\mathcal{U}$ decides the leaf node for each sample based on the comparison results and collects prediction results based on the weights of leaf nodes.
Under such way, we guarantee nodes of the CART model are unable to be accessed by $\mathcal{U}$ and $\mathcal{E}$.
\begin{algorithm}[ht]
  \caption{Secure Prediction for a {CART} (\texttt{SecPred})}
  \label{SecPred}
  \begin{algorithmic}[1]
    \Require
      $\mathcal{S}$ gets a CART $f$ and the thresholds for its nodes, $\{\vartheta_1, \vartheta_2, ..., \vartheta_n\}$; 
      $\mathcal{U}$ gets leaf node weights of $f$.
    \Ensure
      The prediction result.
    
    \For {$1 \leq i \leq n$}
        \State $\mathcal{S}$ computes $\{(\mathcal{S}, \zeta_{\mathcal{S}, e_j})| e_j\in \mathcal{E} \} \gets$\texttt{SS.Share}$(\vartheta_i, t_{\mathcal{E}},$ $ |\mathcal{E}|)$ and sends them to the corresponding edge server.
    
        \For{each $u\in \mathcal{U}$}
            \State Select the feature values $\Omega_{u} = \{\varrho_1, \varrho_2, ...\}$ corresponding to $\vartheta_i$.
            \State Compute $\{(u, \zeta_{u, e_j})| e\in \mathcal{E} \} \gets $ \texttt{SS.Share} $(\varrho \in \Omega_{u}, t_{\mathcal{E}}, |\mathcal{E}|)$ and send to the corresponding edge server.
        \EndFor
        \State $\mathcal{E}$ invokes \texttt{SecCmp}$(\{(e_j, \zeta_{\mathcal{S}, e_j})| e_j\in \mathcal{E} \}, \{(u, \zeta_{u, e_j})|$ $ e_j\in \mathcal{E} \})$ and forwards corresponding results to $u\in \mathcal{U}$.
    \EndFor
    \State Based on the results, $u\in \mathcal{U}$ determines the leaf node and obtains the prediction result.
  \end{algorithmic}
\end{algorithm}

\subsection{Robustness against User Dropout.} \label{sec_UserDrop}
Three possible cases of user dropout in \sysname are discussed as follows. 

\textbf{Case 1:} A user $u_0$ drops out at the \textit{Step 1} or \textit{Step 3} of \textbf{Protocol \ref{FedLearning}}. 
Thus, the edge server $e_0\in\mathcal{E}$ of $u_0$ cannot receive messages from $u_0$ anymore.
In such case, $e_0$ reject the message $u_0$ uploaded in the current round of training.
    
\textbf{Case 2:} A user $u_0$ drops out during the split finding process.
Its edge server $e_0\in\mathcal{E}$ recovers the private mask key of $u_0$ and removes $u_0$ from $\mathcal{U}_0'$ if $u_0$ does not reconnect in the subsequent computation,
that is, the remaining user set $\mathcal{U}_0''\subseteq \mathcal{U}_0'$ and $u_0\in (\mathcal{U}_0'$ $\backslash$ $\mathcal{U}_0'')$.
To recover the private mask key of $u_0$,
    the edge server $e_0$ first collects the random shares of private mask key from at least $t_0$ users, i.e.,  $\zeta_{u_0, v}^{sk}\in \mathcal{R}_v$, $v\in \mathcal{U}_0''$ and $|\mathcal{U}_0''|>t_0$.
Then, $e_0$ extracts the private mask key of $u_0$ through $\langle sk \rangle^{pri}_{u_0}\gets $\texttt{SS.Recon}$(\{(v,$ $ \zeta_{u_0, v}^{sk})\}_{v\in\mathcal{U}''}, t_0)$. 
Finally, $e_0$ retrieves the gradient aggregation result (line 11 and 20, \textbf{Protocol~\ref{SecFind}}) by adding the recomputed values, that is, $\sum_{u\in \mathcal{U}_j'} (\zeta_u - r_u)) + \sum_{u_0 > v} \langle sk \rangle_{u_0, v} - \sum_{u_0 < v} \langle sk \rangle_{u_0, v}$.
Here, $\langle sk \rangle_{u_0, v} = $\texttt{KEY.Agr}$(\langle sk \rangle^{pri}_{u_0}, \langle sk \rangle^{pub}_{v})$.

\textbf{Case 3:} A user $u_0$ drops out at the prediction step.
The edge server $e_0\in\mathcal{E}$ directly ignores the prediction request of $u_0$ and removes $u_0$ from the active users at the next iteration.
\vspace{-0.4cm}

\section{Security Analysis}
\label{SecurityAnalysis}
The FedXGB security depends on three protocols, \texttt{SecFind} and \texttt{SecBoost} and \texttt{SecPred}.
To prove their security, we give the formal definition of security for secret sharing protocol \textit{Definition 2} \cite{ma2019lightweight} and \textit{Theorem 1} from \cite{huang2019lightweight}.

\noindent \textbf{Definition 2.} \textit{We say that a protocol $\pi$ is secure if there exists a probabilistic polynomial-time simulator $\xi$ that can generate a view for the adversary $\mathcal{A}$ in the real world and the view is computationally indistinguishable from its real view.}

\noindent \textbf{Theorem 1.} \textit{The protocol \texttt{SecCmp} is secure in the honest-but-curious security model.} Since the security of \texttt{SecPred} is only related to \texttt{SecCmp}, we omit its security proof which has been given in \cite{huang2019lightweight}. The security of \texttt{SecFind} and \texttt{SecBoost} is proved as follows. 

\noindent \textbf{Theorem 2.} \textit{The protocol \texttt{SecFind} is secure in the honest-but-curious security model.}

\noindent \textit{Proof.}\quad Denote the views of user and edge server as $\mathcal{V}_u = \{view_{u_1}, ...,view_{u_{n}}\}$ and $\mathcal{V}_e = \{view_{e_1}, ..., view_{e_{\theta}}\}$.
From the operation process of \texttt{SecFind}, we can derive 
$view_{e_j} = \{\zeta_u^H, \zeta_u^G, H_j, G_j, \zeta_u^{H, L}, \zeta_u^{G, L}, H_j^L, G_j^L, r_u, \langle ek \rangle_{u, \mathcal{S}}, \langle ek \rangle_{e_j, \mathcal{S}}\}$,
$view_{u} = \{\zeta_u^H, \zeta_u^G, \zeta_{u,v}^{sk}, \zeta_{u,v}^{r}, r_u, A, \zeta_u^{H, L}, \zeta_u^{G, L}, \langle ek \rangle_{u, \mathcal{S}},$ $ \langle ek \rangle_{u, v}, \langle sk \rangle_{u, v}\}$
and
$view_{\mathcal{S}} = \{H, G, H_j, G_j, A, H_L, H_R,$ $ H_j^L, H_j^R, G_L, \\ G_R, G_j^L,G_j^R, score, \langle ek \rangle_{e_j, \mathcal{S}}\}$, 
where $u\in \mathcal{U}_j$, $v\in \mathcal{U}_j$, $e_j\in \mathcal{E}$ and $u\neq v$. Based on \textit{Theorem 1}, it can be seen that, except the XGBoost parameters, the elements belonging to $view_{u}$, $view_{e_j}$ and $view_{\mathcal{S}}$ are all uniformly random shares. According to Shamir's secret sharing theory~\cite{mceliece1981sharing}, the shares can be simulated by randomly chosen values from $\mathcal{F}_p$. Consequently, there exists a simulator $\xi$ that can generate indistinguishable simulated view from the real view of \texttt{SecFind}. According to \textit{Definition 1}, it is derived that the protocol is secure. 

\noindent \textbf{Theorem 3.} \textit{The protocol \texttt{SecBoost} is secure in the honest-but-curious security model.}

\noindent \textit{Proof.}\quad In the protocol \texttt{SecBoost}, the user and edge server views denoted as $\mathcal{V}_u = \{view_{u_1}, ..., view_{u_{n}}\}$ and $\mathcal{V}_e = \{view_{e_1}, ..., view_{e_{\theta}}\}$. 
From the protocol definition, only parts of users are selected for model training in \texttt{SecBoost}. The views of unselected users are set to be empty. The views of remaining users $\mathcal{U}'\subseteq \mathcal{U}$ are
$view_{u} = \{\langle k \rangle_{u}^{sig}, \langle k \rangle_{u}^{ver}, \langle sk \rangle_{u}^{pri},\\ \langle sk \rangle_{u}^{pub}, \langle ek \rangle_{u}^{pri}, \langle ek \rangle_{u}^{pub}, \sigma_{v}, c_{v, u}, \mathcal{R}_{u},$ $ view_{u}'\}$. 
And for the edge server and the cloud server, their views are 
$view_{e_j} = \{\langle k \rangle_{u}^{ver}, \langle sk \rangle_{e_j}^{pub}, \langle ek \rangle_{e_j}^{pub}, \sigma_{u}, c_{v, u},$ $ view_{e_j}'\}$ and
$view_{\mathcal{S}} = \{\langle k \rangle_{u}^{ver}, \langle sk \rangle_{u}^{pub}, \langle ek \rangle_{u}^{pub}, \sigma_{u}, c_{v, u},$ $ view_{\mathcal{S}}'\}$, 
where $u\in \mathcal{U}_j'$, $v\in \mathcal{U}_j'$, $e_j\in \mathcal{E}$ and $u\neq v$. $view_{u}'$, $view_{e_j}'$ and $view_{\mathcal{S}}'$ are the views generated by \texttt{SecFind}. Except the encryption keys, ciphertext and signature which can be treated as random values, the remaining elements of $view_{u}$, $view_{e_j}$ and $view_{\mathcal{S}}$ are all random shares as mentioned in the security proof of \textit{Theorem 2}. Thus, similarly, we can derive that $view_{u}$, $view_{e_j}$ and $view_{\mathcal{S}}$ are simulatable for the simulator $\xi$, and the simulated views cannot be distinguished within a polynomial time by the adversary. Based on \textit{Definition 1}, \texttt{SecBoost} is proved to be secure.

\noindent \textbf{Lemma 1.} A protocol is perfectly simulatable if all its sub-protocols are perfectly simulatable.

According to universal composibility theory given in \textit{Lemma 1} \cite{bogdanov2008sharemind} and the above proofs, it is concluded that \sysname is simulatable. Based on the formal definition of security in \textit{Definition 2}, \sysname is secure.

\section{Performance Evaluation}
\label{Experiments}
In this section, we first introduce the experiment configuration. Then we analyze the effectiveness and efficiency of \sysname by conducting experiments.
\vspace{-0.4cm}

\subsection{Experiment Configuration}
\textbf{Environment.}
A workstation,
    with an Intel(R) Core(TM) i7-7920HQ CPU @3.10GHz and 64.00GB of RAM,
    is utilized to serve as our central server. 
Ten computers, with an Intel(R) Core(TM) i5-7400 CPU @3.00GHz and 8.00GB of RAM, are set up.
By launching multiple processes,
    each of them simulates at most two edge servers.
We also deploy 30 standard BeagleBone Black development boards to serve as crowdsensing users.
Each of them simulates at most 30 users.
The programs are implemented in C++. 
OpenMP library \cite{muddukrishna2016grain} is used to accelerate the concurrent operations.

\textbf{Dataset.}
Two datasets are collected,
    ADULT\footnote{ADULT: https://www.csie.ntu.edu.tw/~cjlin/libsvmtools/datasets/binary.html} and MNIST\footnote{MNIST: http://yann.lecun.com/exdb/mnist/}.
ADULT is for adult income prediction, 
    which has 123 features, and provides 32k instances for training data, 16k instances for testing. 
MNIST is for handwriting digit classification, 
    which has 784 features, and divides the instances into 60k for training and 10k for testing. 
Both are commonly used databases to evaluate machine learning model performance.

\textbf{Setup.}
Parameters in \sysname are set up as, 
    step size $\eta = 0.3$, minimum loss reduction $\gamma = 0.1$, regulation rate $\lambda = 1.0$, user number $n = 300$, maximum tree depth $d_{max} = 3$, and edge server number $\theta = 10$. 
We use Elliptic-Curve Diffie-Hellman~\cite{doerner2018secure} over the NIST P-256 curve, composed with a SHA-256 hash, to fulfill key agreement,.
Authenticated encryption is operated by 128-bit {AES-GCM}~\cite{bellare2016multi}.
Given each dataset,
    the instances are averagely assigned to each user with no overlap. 
User dropout is assumed to occur every 10 rounds of boosting in our experiment.
That is,
    0\%, 10\%, 20\%, 30\% of users are randomly selected to be disconnected at each $10^{th}$ round of training.
Meanwhile, the same number of replacements are rearranged to substitute the lost users.
\vspace{-0.3cm}

\subsection{Effectiveness Analysis}
\begin{figure}[htbp]
\centering
\subfigure[Accuracy with different user dropout rates for ADULT.]{
\begin{minipage}[t]{0.45\linewidth}
\centering
\includegraphics[scale=0.18]{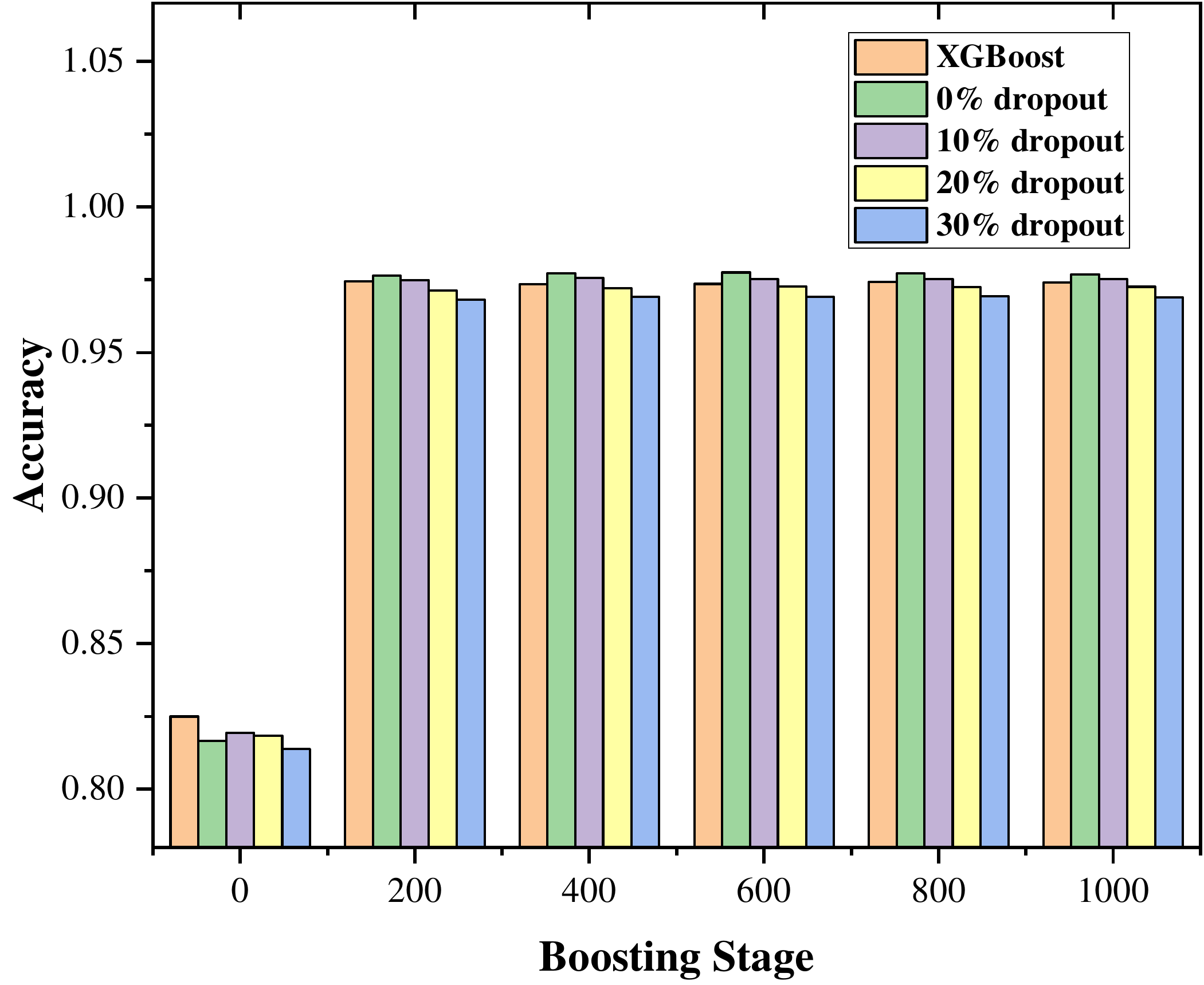}
\label{Accuracy_Adult}
\end{minipage}
}%
\hfill
\subfigure[Loss with different user dropout rates for ADULT.]{
\begin{minipage}[t]{0.45\linewidth}
\centering
\includegraphics[scale=0.18]{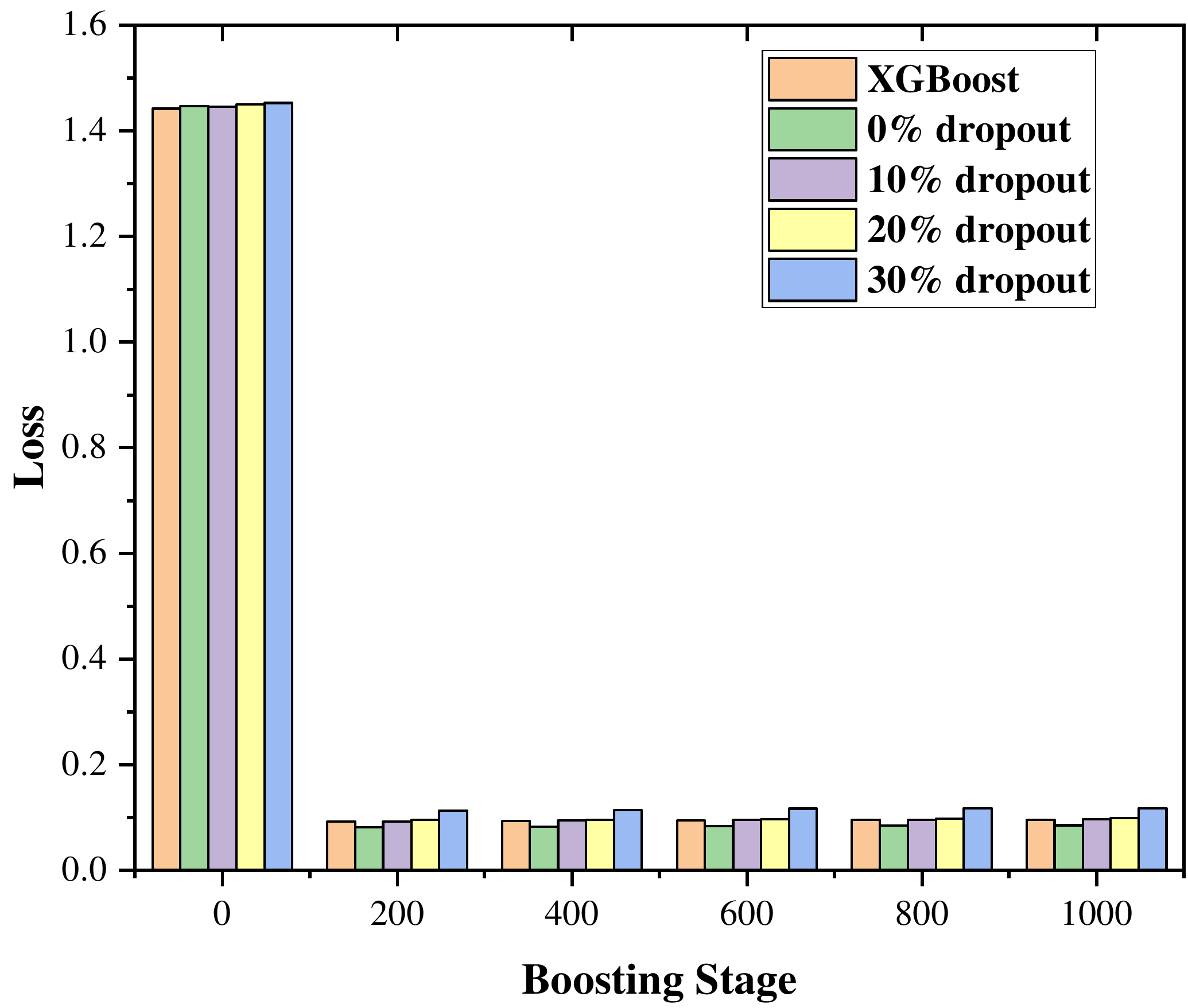}
\label{Loss_Adult}
\end{minipage}
}%
\hfill
\subfigure[Accuracy with different user dropout rates for MNIST.]{
\begin{minipage}[t]{0.45\linewidth}
\centering
\includegraphics[scale=0.175]{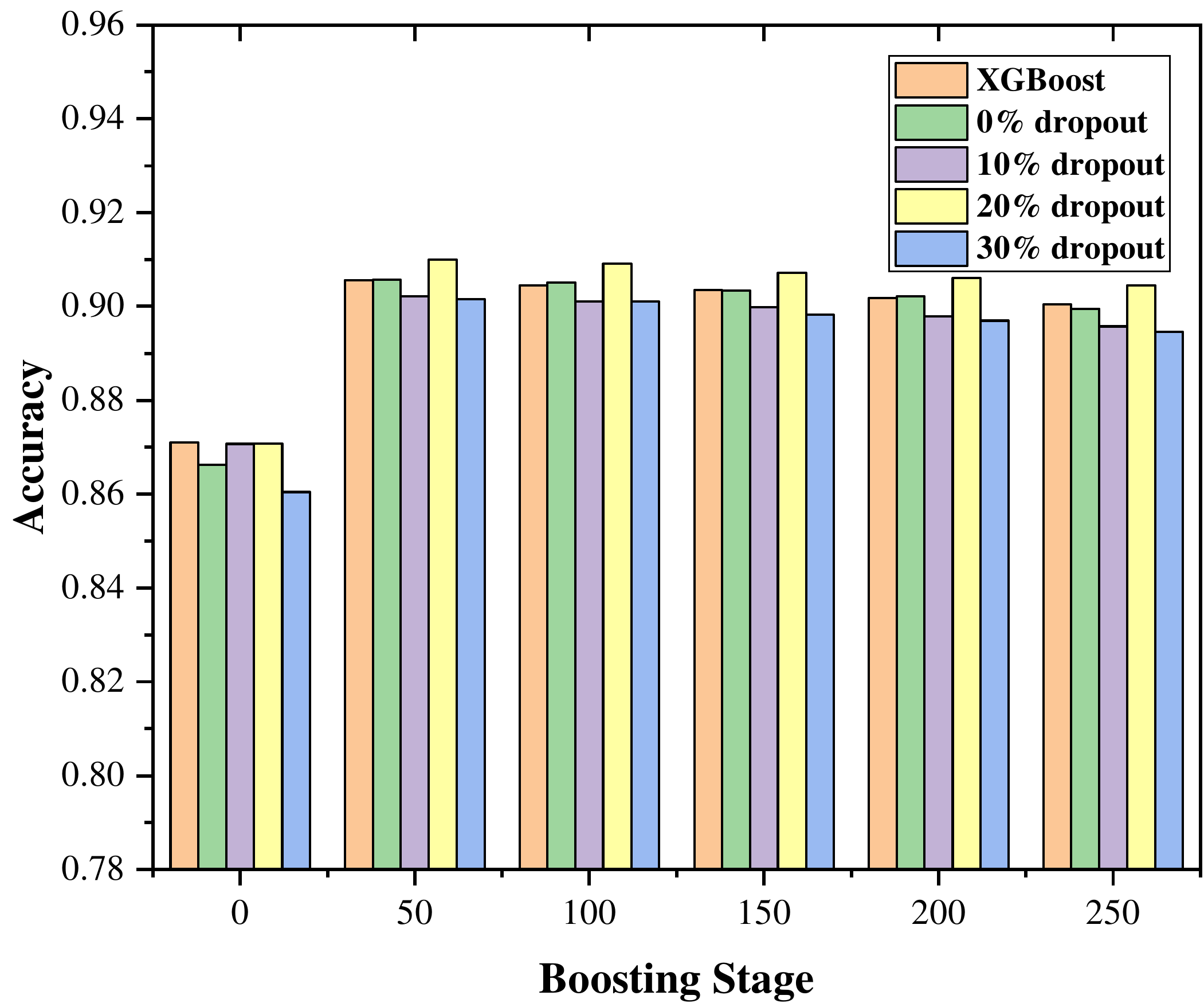}
\label{Accuracy_MNIST}
\end{minipage}
}%
\hfill
\subfigure[Loss with different user dropout rates for MNIST.]{
\begin{minipage}[t]{0.45\linewidth}
\centering
\includegraphics[scale=0.18]{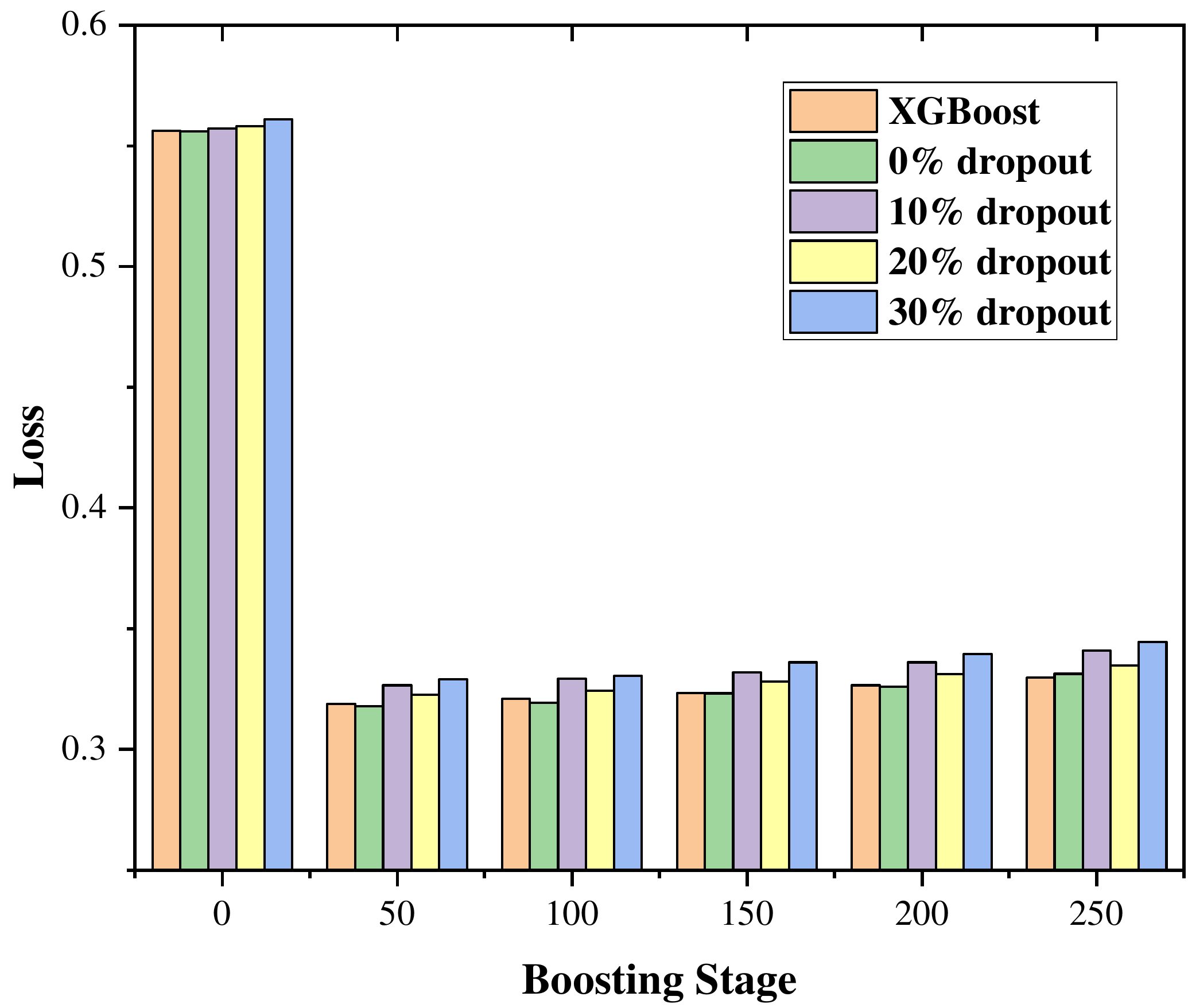}
\label{Loss_MNIST}
\end{minipage}
}%
\caption{Accuracy and loss for each boosting stage in \sysname for ADULT and MNIST}
\label{Auc_Loss}
\end{figure}

To assess the effectiveness of \sysname, we compute its classification accuracy and loss under the two datasets. 
The loss functions utilize in the experiments are the logistic regression for ADULT and the softmax for MNIST.
We evaluate the accuracy and loss for each boosting stage in \sysname, shown in Fig.~\ref{Auc_Loss}.
Fig.~\ref{Accuracy_MNIST} and Fig.~\ref{Loss_MNIST} respectively show the accuracy and loss of MNIST, and Fig.~\ref{Accuracy_Adult} and Fig.~\ref{Loss_Adult} reveal the result of ADULT.
When boosting stage is determined, \sysname has only a loss of less than 1\% compared with the original XGBoost.
\sysname acquires small fault tolerance with the user dropout rate ranged from 0\% to 30\%.

\subsection{Efficiency Analysis}
\subsubsection{Theoretical Analysis}
To evaluate the efficiency of \sysname, we first perform the theoretical analysis of computation cost for \func{SecBoost}, \func{SecFind} and \func{SecPred}. 

Let $|\mathcal{D}|$ denote the number of training instances.
The computation costs of each user, 
    each edge server and the central server for \func{SecBoost} are $\mathcal{O}(n/\theta + (n/\theta) \cdot \delta d_{max} |\mathcal{D}|)$, $\mathcal{O}(n/\theta + (n/\theta)^2\cdot \delta d_{max} |\mathcal{D}|)$ and $\mathcal{O}(n + \delta \theta d_{max})$.
As shown in \textbf{Protocol~\ref{FedLearning}}, 
    \func{SecBoost} has four steps.
Since the setup stage can be operated offline,
    its computation and communications cost are ignored. 
The remaining three steps are divided into two parts.
One part contains the second and third steps. 
Each user executes $2(n/\theta)$ key agreements, signature and encryption operations, which take $\mathcal{O}(n/\theta)$ time.
Each edge server executes $n/\theta$ signature operations, which also take $\mathcal{O}(n/\theta)$ time.
The central server executes $n$ signature operations, which take $\mathcal{O}(n)$ time.
The other part is composed of $d_{max}$ invocations of \func{SecFind}.
And for \func{SecFind}, the gradient aggregation is operated for $\delta$ times, which takes $\mathcal{O}((n/\theta) \cdot \delta |\mathcal{D}|)$, $\mathcal{O}((n/\theta)^2 \cdot \delta |\mathcal{D}|)$ and $\mathcal{O}(\delta \theta)$ time for each user, each edge server and the central server.
As for the \func{SecPred}, it invokes \func{SecCmp} for $d_{max}$ times, which takes  $\mathcal{O}(|\mathcal{D}|)$, $\mathcal{O}(d_{max}|\mathcal{D}|)$ and $\mathcal{O}(d_{max})$ time.

\subsubsection{Experiment Results}
\begin{figure*}[htbp]
\centering
\subfigure[Runtime per user with different numbers of users.]{
\begin{minipage}[t]{0.21\linewidth}
\centering
\includegraphics[scale=0.175]{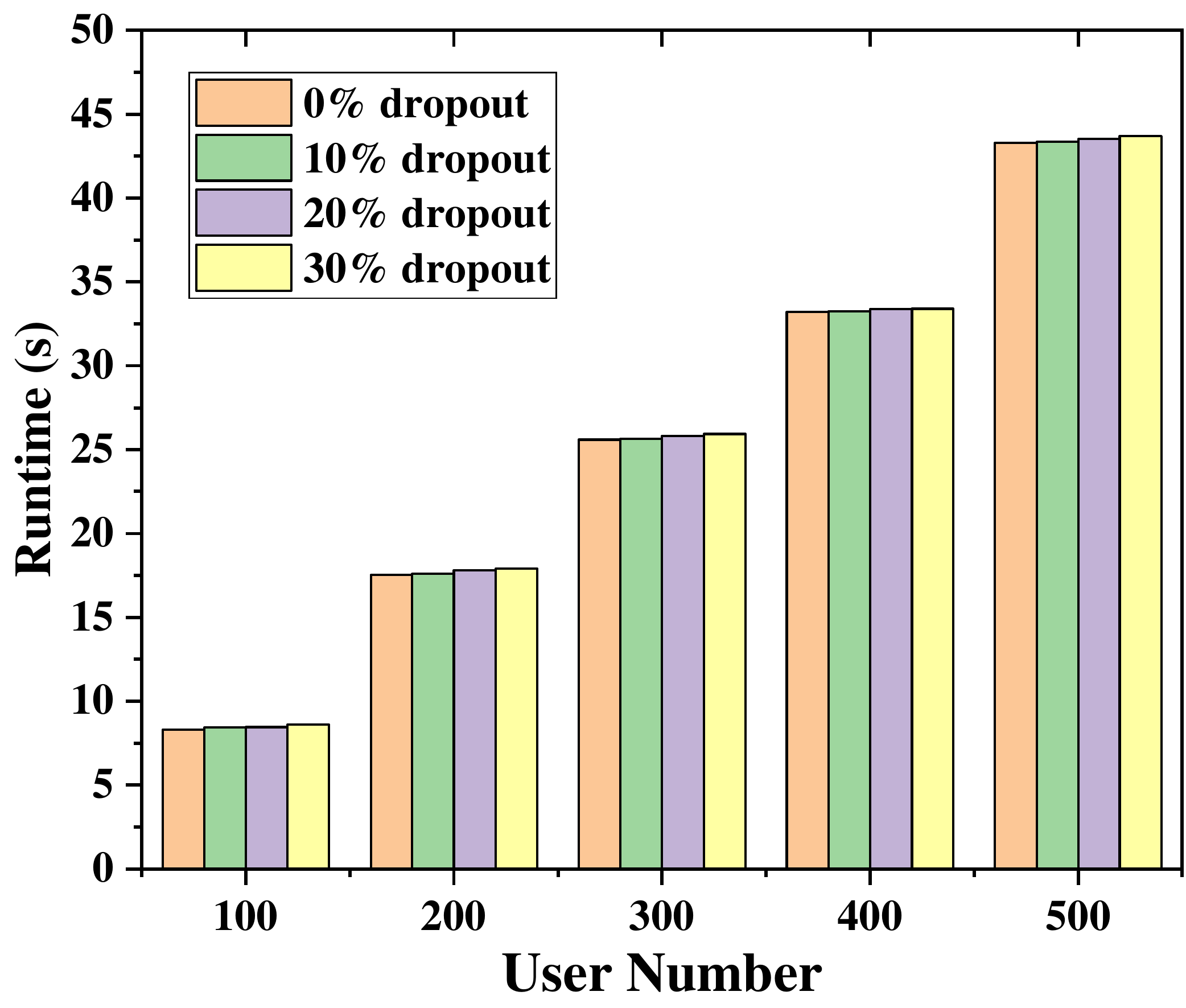}
\label{runtime_user_user}
\end{minipage}
}%
\hfill
\subfigure[Communication overhead per user with different numbers of users.]{
\begin{minipage}[t]{0.21\linewidth}
\centering
\includegraphics[scale=0.175]{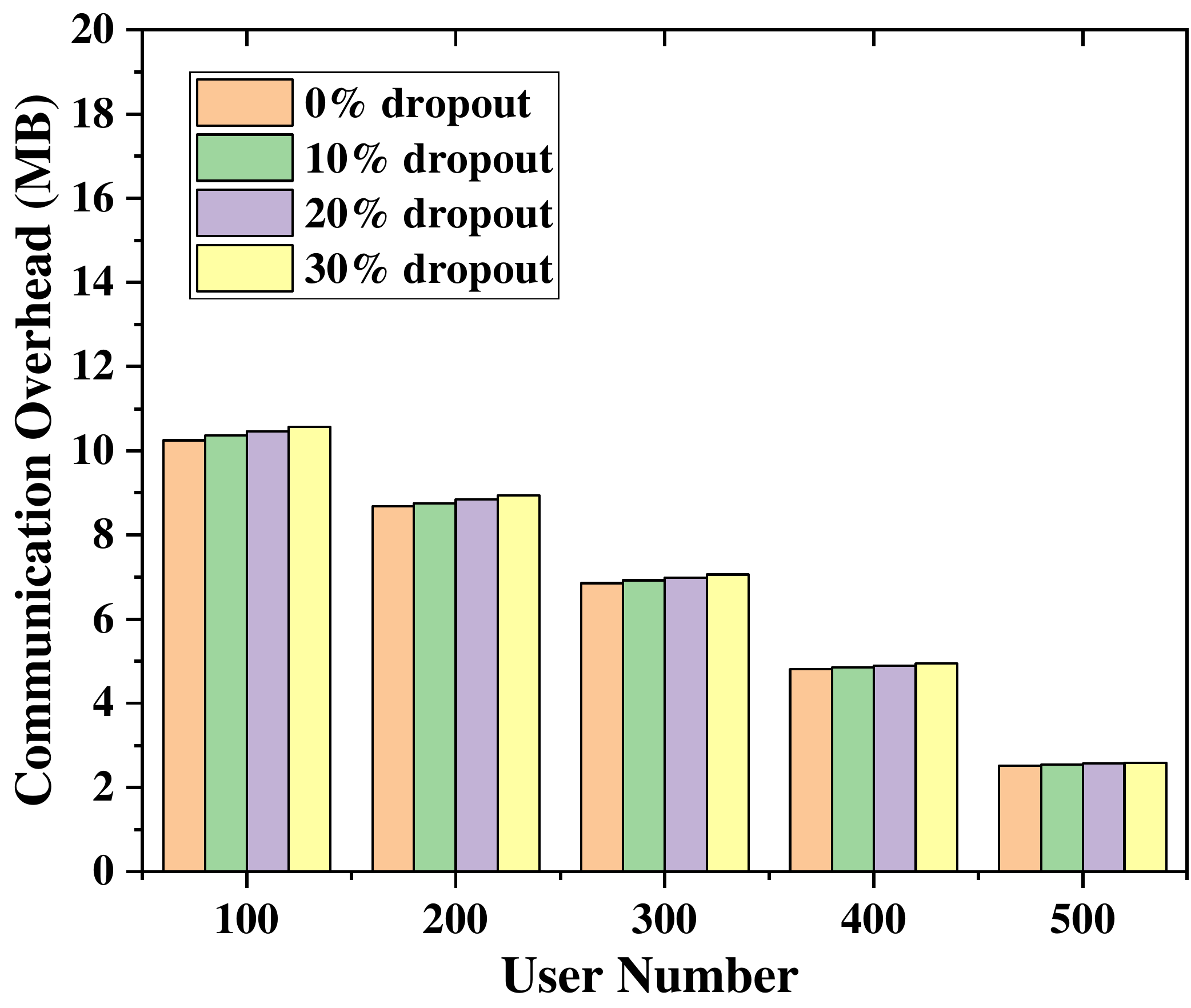}
\label{comm_user_user}
\end{minipage}
}%
\hfill
\subfigure[Runtime per edge server with different numbers of users.]{
\begin{minipage}[t]{0.21\linewidth}
\centering
\includegraphics[scale=0.175]{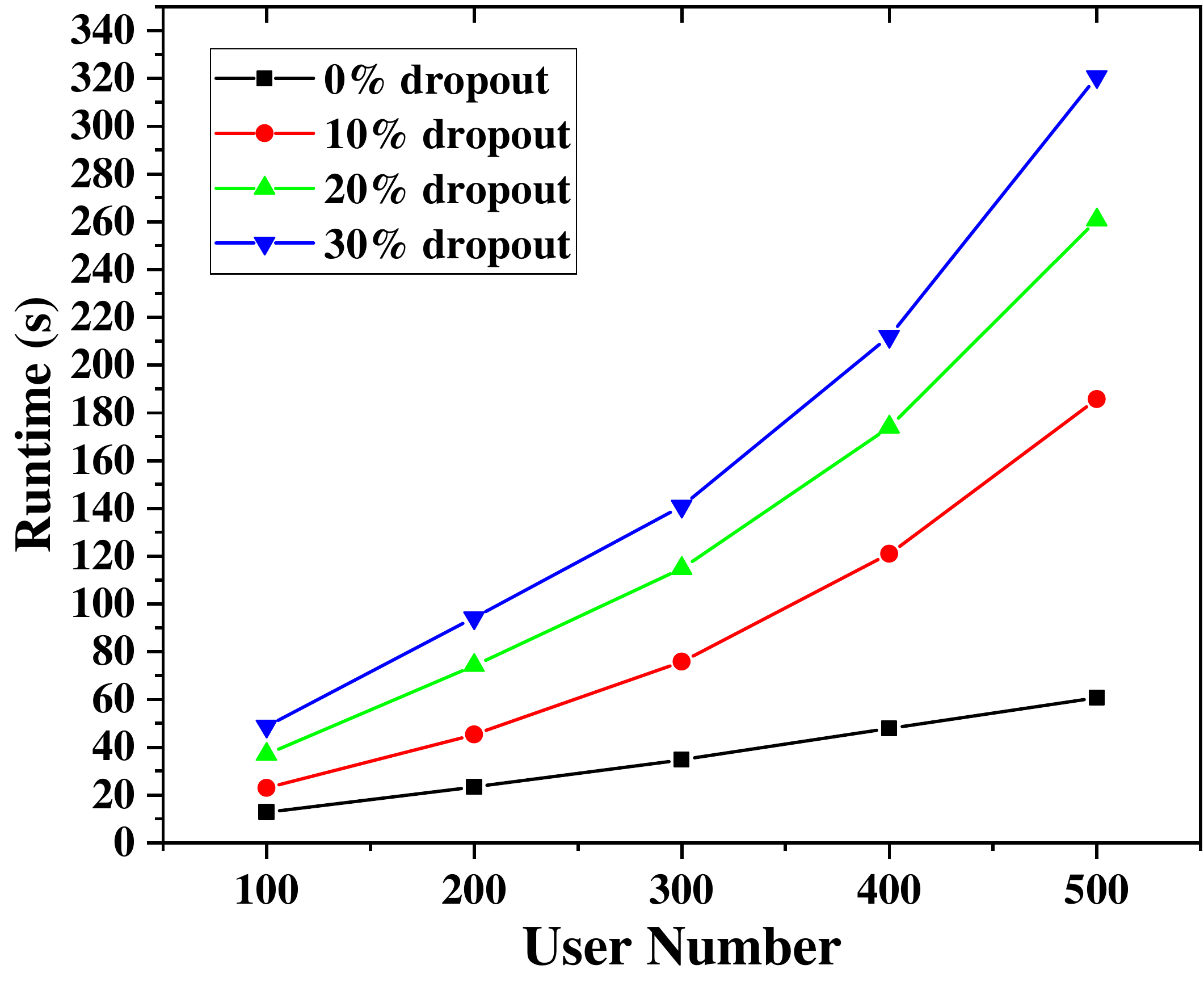}
\label{runtime_user_edge}
\end{minipage}
}%
\hfill
\subfigure[Communication overhead per edge server with different numbers of users.]{
\begin{minipage}[t]{0.21\linewidth}
\centering
\includegraphics[scale=0.175]{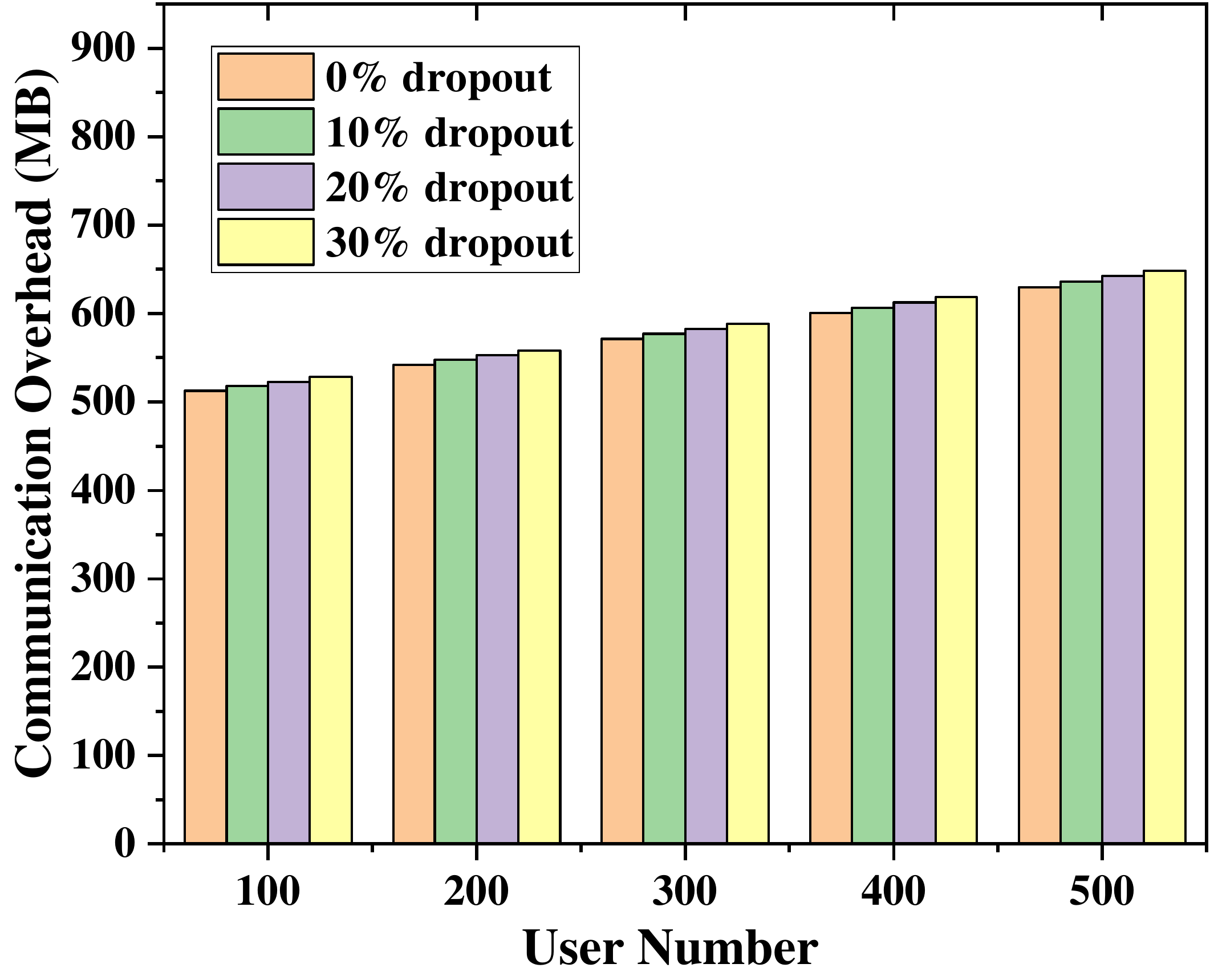}
\label{comm_user_edge}
\end{minipage}
}%
\hfill
\subfigure[Runtime for the central server with different numbers of users.]{
\begin{minipage}[t]{0.21\linewidth}
\centering
\includegraphics[scale=0.175]{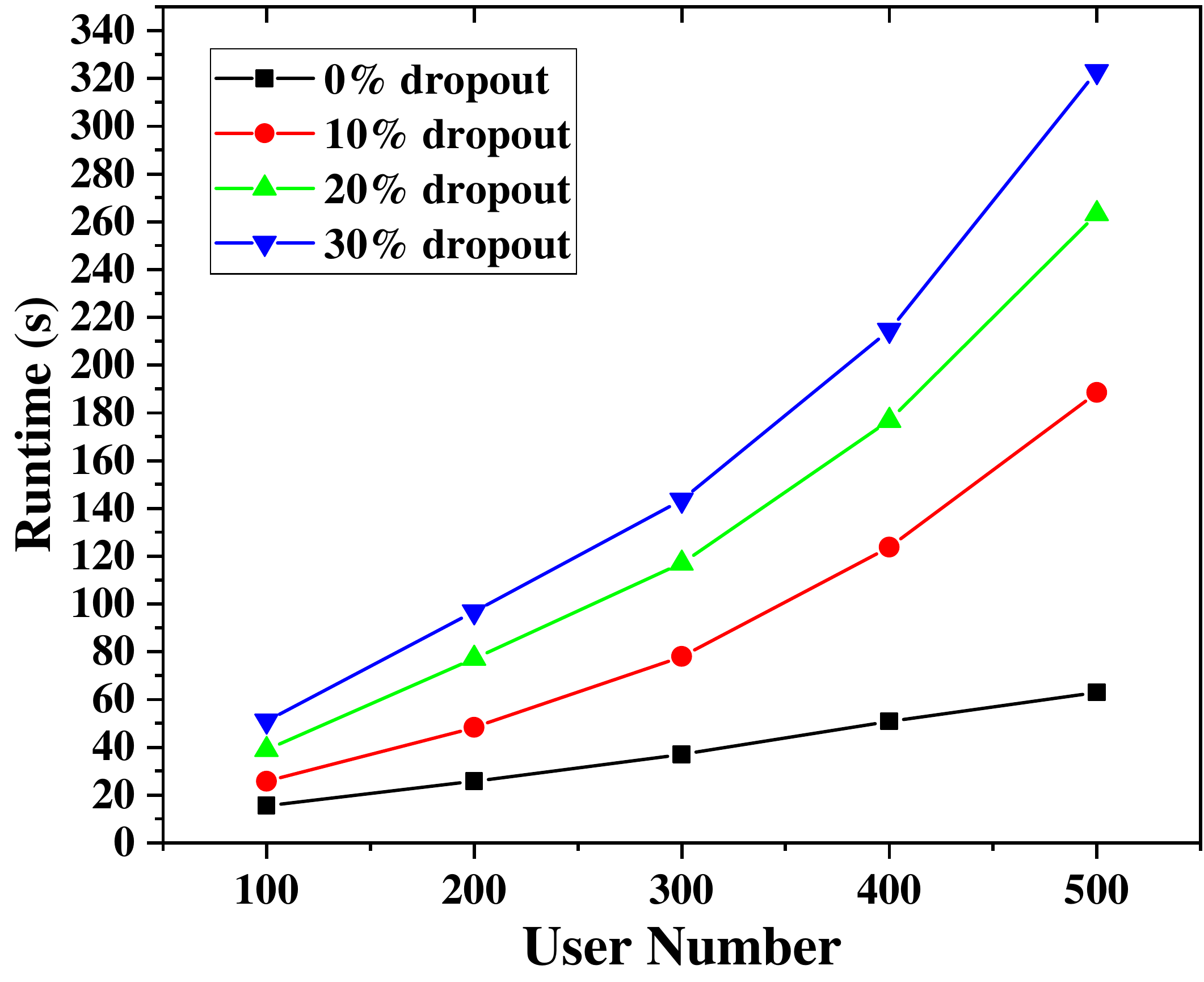}
\label{runtime_user_central}
\end{minipage}
}%
\hfill
\subfigure[Runtime per user with different numbers of edge servers.]{
\begin{minipage}[t]{0.21\linewidth}
\centering
\includegraphics[scale=0.175]{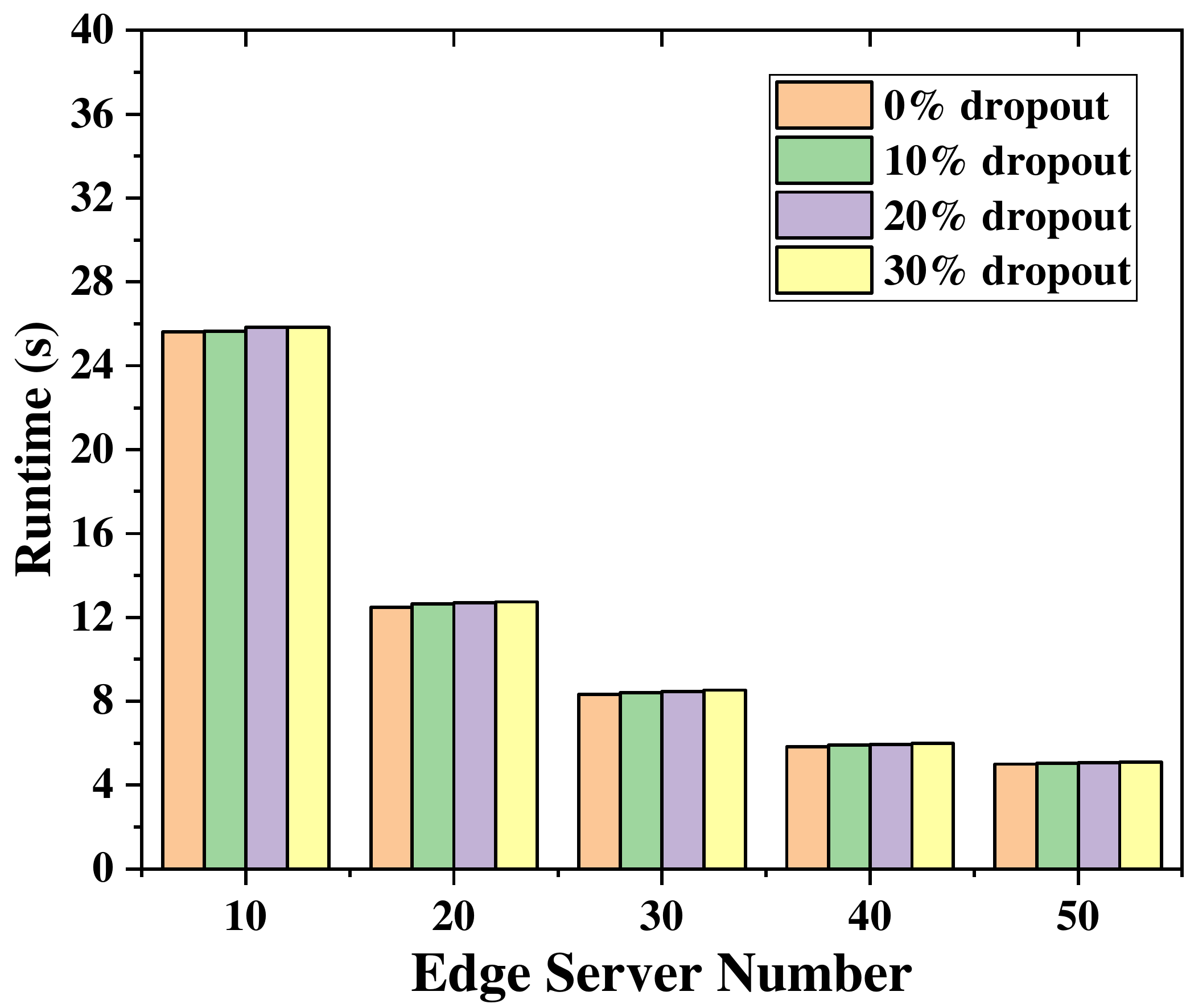}
\label{runtime_edge_user}
\end{minipage}
}%
\hfill
\subfigure[Runtime per edge server with different numbers of edge servers.]{
\begin{minipage}[t]{0.21\linewidth}
\centering
\includegraphics[scale=0.175]{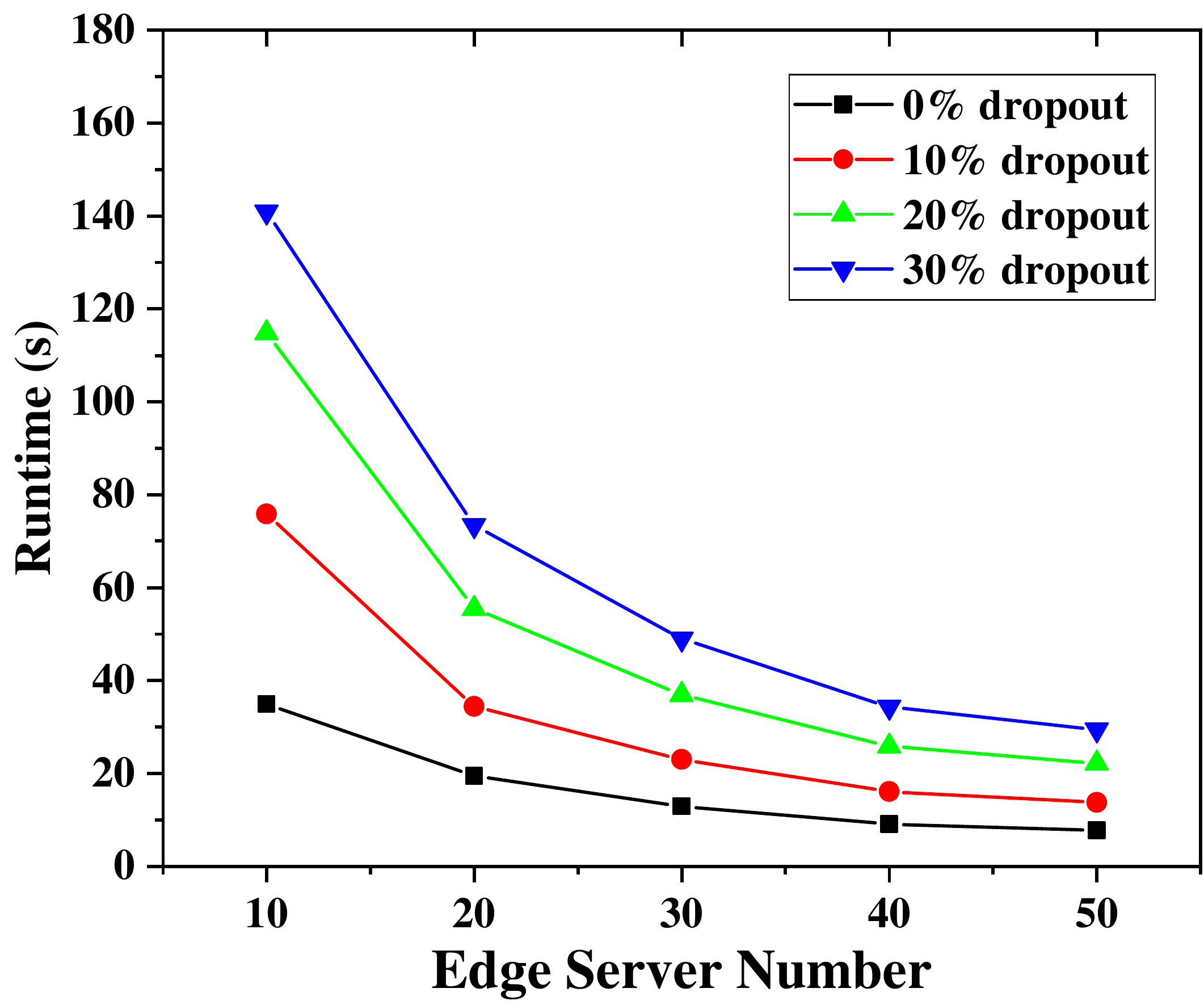}
\label{runtime_edge_edge}
\end{minipage}
}%
\hfill
\subfigure[Communication overhead per edge server with different numbers of edge servers.]{
\begin{minipage}[t]{0.21\linewidth}
\centering
\includegraphics[scale=0.175]{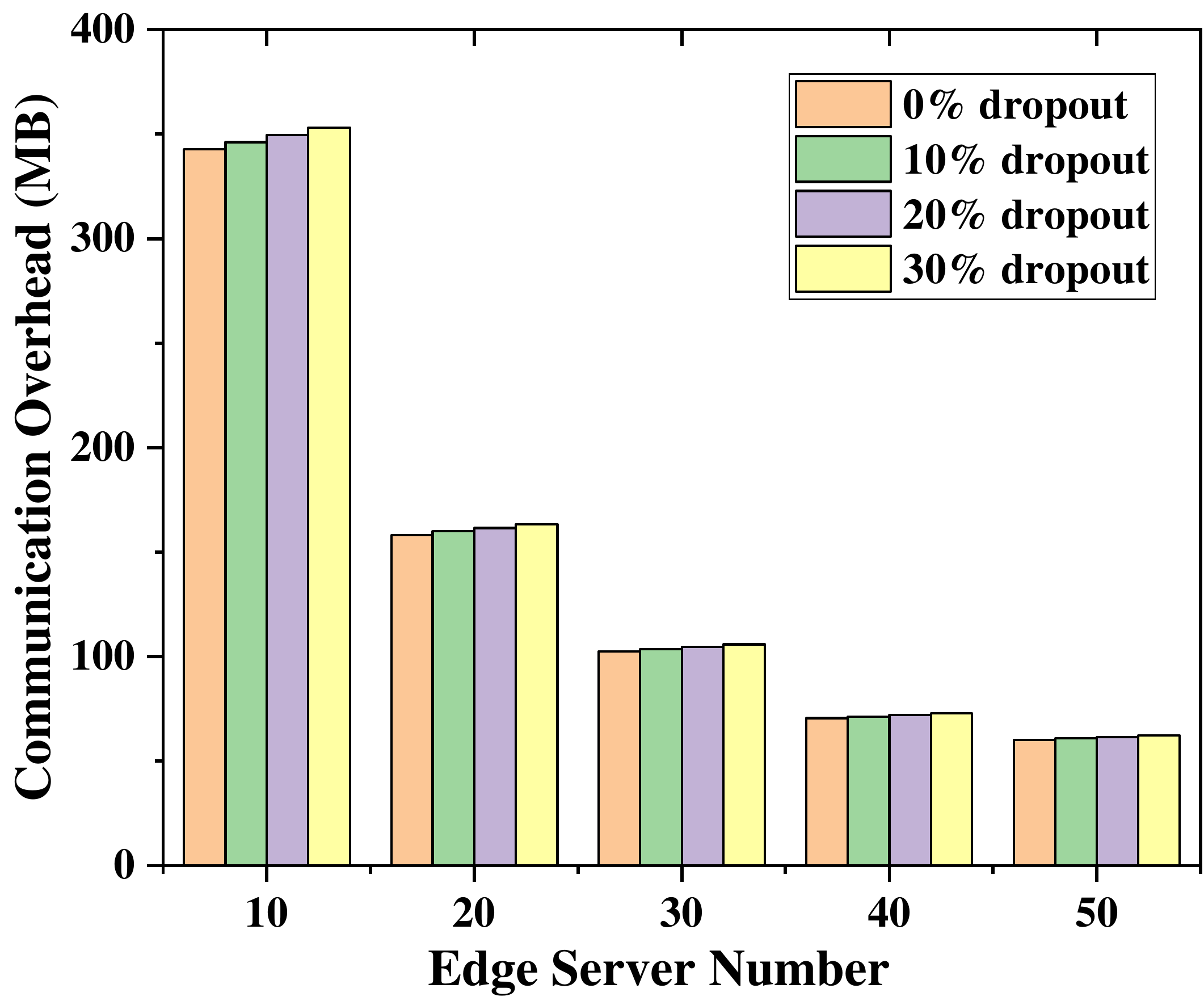}
\label{comm_edge_edge}
\end{minipage}
}
\caption{Runtime and communication overhead with different numbers of users and edge servers.}
\label{efficiency_feature_sample}
\end{figure*}

To further evaluate the efficiency of \sysname, we experiment with the runtime and communication overhead under different numbers of users and edge servers as shown in Fig.~\ref{efficiency_feature_sample}. 
In the experiments, we set $|D|=$ 50K and $\delta = 100$.

\textbf{Number of Users.}
When the involved users increase,
    the runtime for each user grows linearly, and inversely, the communication overhead for each user decreases, 
    shown in Fig.~\ref{runtime_user_user} and Fig.~\ref{comm_user_user}, respectively.
The linear growth of the runtime is caused by the incremental cost of user selection and key shares collection steps.
And due to the less samples distributed to each user, the communication overhead for each user decreases.
The user dropout rate barely influences the runtime because the correlated active user only need to transmit one secret sharing for the private mask key reconstruction.
Considering the impact of the incremental user number performed on each edge server,
    the runtime for each edge server follows the quadratic growth, 
    described in Fig.~\ref{runtime_user_edge}.
The private mask key recovery for dropped users has the main effect on the increase of the runtime cost.
Nonetheless,
    the communication overhead is barely influenced because only a little overhead increment is caused for the key shares collection stage of \func{SecBoost}.
The higher user dropout causes obvious time increment to reconstruct lost data via the time-consuming Lagrange polynomials.    
Specially, the central server deploys less computation tasks than edge server, but has more runtime as illustrated in Fig.\ref{runtime_user_central}. 
The phenomenon is due to the fact that the central server has to wait for collecting every edge server's response to continue the subsequent computation. 
The communication overhead plots about central server are omitted, because, for central server, its communication overhead is just the edge server number multiplied the difference between the edge server overhead and the user overhead. 

\textbf{Number of Edge Servers.}
When the involved edge servers increase,
    the runtime cost for each user decreases, illustrated in Fig.~\ref{runtime_edge_user}.
Because the number of users in each domain managed by each edge server reduces,
    the computational cost of secret sharing also becomes less for each user.
Similarly,
    the runtime cost of each edge server decreases, Fig.~\ref{runtime_edge_edge} while the computation of secret sharing assigned on each edge server reduces.
As more edge servers are involved for computation, 
    the communication overhead of each server decreases, shown in Fig.~\ref{comm_edge_edge}.
For each user,
    the communication overhead does not have obvious change because the assigned instances are static.
And the cost of central server performs similar to Fig.~\ref{runtime_user_central} with $300$ users.
Due to the space limitation, we omit these two plots in this paper.
\vspace{-0.2cm}
\renewcommand\arraystretch{1.5}
\begin{table}[!htbp]
\centering
\caption{Protocol runtime for different Stages}
\begin{tabular}{c|c|c|c|c|c}

\hline
\multicolumn{1}{c|}{\multirow{3}{*}{Stage}}     & \multicolumn{4}{c}{RunTime (s)}\\

\cline{2-5}
& \multicolumn{3}{c|}{Our \sysname}    &  \multicolumn{1}{c}{\multirow{2}{*}{SecureBoost \cite{cheng2019secureboost}}}\\

\cline{2-4}
& $\mathcal{U}$        & $\mathcal{E}$   &    $\mathcal{S}$   \\

\hline
User Selection      & 0.285     & 1.112        & \multicolumn{1}{c|}{3.288}          & \multicolumn{1}{c}{\multirow{2}{*}{N.A.}}\\

\cline{1-4}
Mask Collection     & 1.333     & 1.458        & \multicolumn{1}{c|}{N.A.}             \\

\hline
Boosting            & 18.802    & 23.308       & \multicolumn{1}{c|}{26.863}         & \multicolumn{1}{c}{\multirow{2}{*}{46.25}}\\

\cline{1-4}
Prediction   & 5.182     & 5.987        & \multicolumn{1}{c|}{6.961}            \\

\hline
Total               & 25.602    & 31.865       & \multicolumn{1}{c|}{37.112}         & \multicolumn{1}{c}{46.25}\\

\hline
\end{tabular}
\label{table_overhead}
\end{table}

In Table~\ref{table_overhead}, 
    we list the runtime cost of different stages in \sysname.
It indicates that the main overhead in \sysname is caused by the boosting stage, 
    namely, the optimal split finding algorithm,
    because numerous loop operations are proceeded.
We also compare the runtime between \sysname and the only existing privacy-preserving XGBoost architecture, SecureBoost, proposed in \cite{cheng2019secureboost}.
SecureBoost deploys the homomorphic encryption algorithm (HE) technology to protect the sub-aggregation of gradients.
Since HE is still a time-consuming technique for multi-party computation \cite{aslett2015review}, SecureBoost takes more time than \sysname to handle the same size instances. 
And different from \sysname, SecureBoost is specially designed for the vertically split distributed data (i.e. the data are split in the feature dimension).
The setting limits its suitable application situation, because for most circumstances of mobile crowdsensing applications, each user independently forms a dataset with all features, like the individual income condition (i.e. the data are horizontally split).
Additionally, the user dropout condition and the model privacy leakage problem are not considered in SecureBoost.
\vspace{-0.2cm}

\subsection{Defense Against User Data Reconstruction Attack}\label{sec_UDR}
\begin{figure}[ht!]
\centering
\includegraphics[scale=1.22]{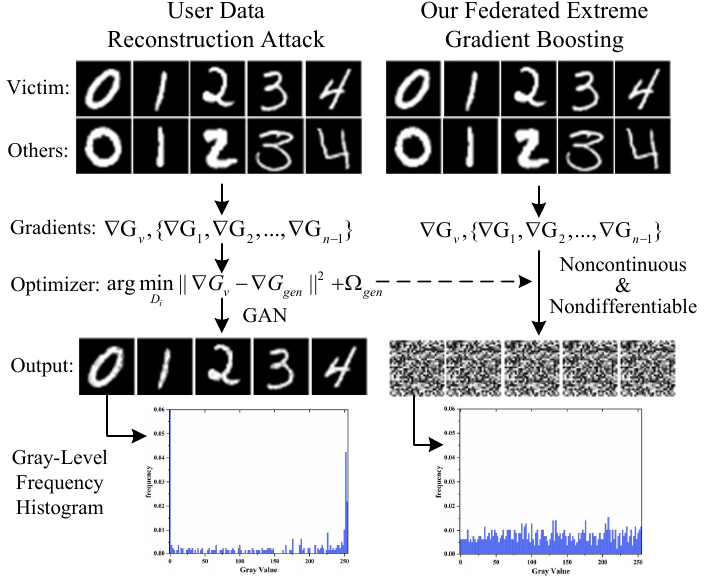}
\caption{Security of FedXGB against user data reconstruction attack}
\label{GAN_Attack}
\end{figure}

GAN based reconstruction attack~\cite{wang2019beyond} is one of the most common and effective attacks against federated learning. 
Based on GAN,
    the attack reconstructs user data by solving an optimization problem.
However,
    \sysname is protected against such GAN-based attack due to the tree structure we choose.
In order to validate how well \sysname is protected,
    we conduct two experiments by launching the user data reconstruction (UDR) attack against the original federated learning approach~\cite{mcmahan2016communication} and \sysname.
In the experiment, MNIST is used, shown in Fig.~\ref{GAN_Attack}.

The left column of Fig.~\ref{GAN_Attack} illustrates that the federated learning approach is attacked successfully.
The attacker (i.e., the central server), $\mathcal{S}$, first collects the gradient aggregations uploaded by the specific victim $\nabla G_v$ and other users $\nabla G_i$, $1 \leq i \leq n - 1$.
Based on $\nabla G_v$ and $\nabla G_i$, 
    the attacker derives the representatives $X_i$ of the victim by solving the optimization problem $Op = \arg\min ||\nabla G_v - \nabla G_{gen}||^2 + \Omega_{gen}$, 
    where $\Omega_{gen}$ is a regularization item and $\nabla G_{gen}$ is the gradient of $X_i$.
Given $X_i$,
    GAN outputs almost identical images.

The right column of Fig.~\ref{GAN_Attack} presents the failed UDR attack launched on \sysname.
Suppose that  $\nabla G_i$ is the gradient aggregation obtained by an malicious edge server $e_j$.
Because the CART model partitions the input space into discrete regions,
    $e_j$ is unable to solve the optimization problem $Op$.
The optimizer can only advance towards random directions and output images that look like random noises.
The gray-level frequency histograms in the last row of Fig.\ref{GAN_Attack} further illustrate that, for \sysname, UDR can hardly fit the features of original images.
\vspace{-0.2cm}

\section{Related Work}
\label{RelatedWork}
Most of the existing privacy-preserving works for machine learning are data driven and based on traditional cryptographic algorithms. For example, Wang \textit{et al.} \cite{wang2017learning} proposed a privacy-preserving data mining model learning scheme for canonical correlation analysis in cross-media retrieval system garbled circuit. Ma \textit{et al.} \cite{ma2019lightweight} proposed a lightweight ensemble classification learning framework for the universal face recognition system by exploiting additive secret sharing. Considering the wide applications of gradient boosting decision tree (GDBT) in data mining, Zhao \textit{et al.} \cite{zhao2018inprivate} utilized the differential privacy technology to implement two novel privacy-preserving schemes for classification and regression tasks. Towards the patient's medical data privacy protection in e-Health system, Liu in \cite{liu2019privacy} advocated a homomorphic encryption based scheme to implement privacy-preserving reinforcement learning scheme for patient-centric dynamic treatment regimes. Because of data security driven, the above four types of privacy-preserving schemes still have to upload encrypted user data to central server and cause massive communication overheads. 

Therefore, the federated learning concept was proposed \cite{mcmahan2016communication}. Up to now, there were only a few works that adapted the architecture to propose practical schemes for applications \cite{tran2019federated}. And most existing federated learning schemes still concentrated on the stochastic gradient descent (SGD) based models. For example, considering the limited bandwidth, precious storage and imperative privacy problem in Internet of Things (IoT) environment, Wang \textit{et al.}~\cite{wang2018edge} provided a SGD based federated machine learning architecture based on the edge nodes. For the privacy-preserving machine learning model training in smart vehicles, Sumudu \textit{et al.} \cite{samarakoon2018federated} proposed a federated learning based novel joint transmit power and resource allocation approach. And to avoid the adversary to analyze the hidden information about user private data from the uploaded gradient values, cryptographic methods were then added to the original federated learning scheme for protecting gradients. Keith \textit{et al.} \cite{bonawitz2017practical} designed a universal and practical model aggregation scheme for mobile devices with secret sharing technology. In \cite{nock2018entity}, Richard \textit{et al.} utilized the homomorphic encryption to protect the uploaded gradients and designed an entity resolution and federated learning framework.

\section{Conclusion} 
\label{Conclusion}
In this paper, we proposed a privacy-preserving federated learning architecture (\sysname) for the training of extreme gradient boosting model (XGBoost) in crowdsensing applications. 
For securely building classification and regression forest of XGBoost, we designed a series of secure protocols based on the secret sharing technique. 
The protocols guarantee that the privacy of user data, learning gradients and model parameters are simultaneously preserved during the model training process of XGBoost. 
Moreover, we conducted numerous experiments to evaluate the effectiveness and efficiency of \sysname. 
Experiment results showed that \sysname is able to support massive crowdsensing users working together to efficiently train a high-performance XGBoost model without data privacy leakage.



\end{document}